\documentclass[twocolumn]{aastex62}

\usepackage{lineno}

\defcitealias{K15}{K15}
\defcitealias{Guo2013PhotoCat}{G13}
\defcitealias{Simard2011BulgeDiskDecompCatalog}{S11}
\defcitealias{Meert2015SDSSBulgeDiskDecompCatalog}{M15}

\begin{document}

\title{\MakeUppercase{Assessment of SDSS-derived Galaxy Morphologies using HST imaging}}

\author{Chandler Osborne}
\affil{Department of Astronomy, Indiana University, Bloomington, IN, 47408}

\author[0000-0003-2342-7501]{Samir Salim}
\affil{Department of Astronomy, Indiana University, Bloomington, IN, 47408}

\begin{abstract}

The Sloan Digital Sky Survey (SDSS) was foundational to the study of galaxy evolution, having revealed the bimodality of galaxies and the relationship between their structure and star-forming activity. However, ground-based optical surveys like SDSS are limited in resolution and depth which may lead to biases or poor quality in the derived morphological properties, potentially impacting our understanding of how and why galaxies cease their star formation (quench). We use archival HST imaging of $\sim2,000$ SDSS objects to assess the reliability of SDSS-derived morphologies, taking advantage of both SDSS statistical samples and of HST's superior resolution and sensitivity. Single S\'ersic fitting and bulge-disk decomposition is performed on HST images for direct comparison with SDSS results. Of the three catalogs of SDSS-derived morphologies considered, none are significantly more accurate than the others. For disk-dominated galaxies ($n<2.5$), global S\'ersic indices ($n$) from \citet[][M15]{Meert2015SDSSBulgeDiskDecompCatalog} are preferred. For bulge-dominated galaxies ($n>2.5$), \citet[][S11]{Simard2011BulgeDiskDecompCatalog} and M15 overestimate $n$ by $\sim20\%$, and NYU-derived global $n$ are preferred. Global $R_{\text{eff}}$ from \citet[][S11]{Simard2011BulgeDiskDecompCatalog} are preferred, but overestimate $R_{\text{eff}}$ for the largest galaxies by $0.1$ dex. SDSS-derived single-component parameters are generally significantly more robust than SDSS-derived two-component parameters. The bulge S\'ersic index ($n_{\text{bulge}}$) cannot be reliably constrained from SDSS imaging. The bulge-to-total (B/T) ratio can be reliably inferred from SDSS for galaxies with SDSS B/T $\lesssim0.6$ provided that $n_{\text{bulge}}=4$ is enforced. The difference in global $n$ between HST and SDSS depends strongly on B/T; an empirical correction based only on it accounts for most of the systematics in global $n$.

\end{abstract}

\section{Introduction}


The advent of large-scale optical surveys like the Sloan Digital Sky Survey \citep[SDSS;][]{YorkSDSSTechnical, Strauss2002SDSSMainGalaxySampleSpecSelection} marked a paradigm shift in the study of galaxy evolution. With uniform spectroscopy out to $z\sim0.3$ and a huge sample size ($10^6$), SDSS revealed the bimodality of the galaxy population, wherein star-forming and passive galaxies occupy distinct regions in optical color space \citep{Strateva2001ColorSepOfGalaxiesInSDSS, Kauffmann2003GalaxyBimodalitySDSS, Baldry2004SDSSGalaxyCMDOpticalBimodality}. The Galaxy Evolution Explorer (GALEX) UV survey additionally revealed the `green valley' which consists of galaxies with intermediate levels of SF \citep{Wyder2007UVOpticalCMDBasicProps}. 


The galaxy optical color bimodality also extends to morphology; optically blue galaxies are preferentially disky late-types whereas optically red galaxies are preferentially early-type spheroids and bulge-dominated disks \citep[e.g.,][]{Wuyts2011StructureEvolution}. The current picture of galaxy evolution places galaxies in a `grow-and-quench' framework where galaxies form as disks and grow their mass in a quasi-equilibrium state maintained by the competing effects of gas accretion and feedback, wherein episodes of star formation (SF) fuel central mass growth and can lead to temporary suppression of SF \citep{Zolotov2015Compaction, Tacchella2016Compaction&ConfinementInTheSFMS}. A permanent shut off of SF, or quenching, can occur once some threshold is crossed, seemingly related to the formation of a central bulge \citep{Cheung2012QuenchingAndInnerStructureAEGIS, Barro2014CompactHighzProgenitorsOfCompactQuiescGals, Lang2014BulgeGrowth&Quenching, Tacchella2016HighzCompaction&Quenching, Osborne2020CANDELSmeetsGSWLC, Chen2020QuenchingContestHalosAndBlackHoles}.

The relative importance of different quenching mechanisms can be established with an accurate census of galaxy morphology. So-called `hallmark' merger signatures such as loops, bridges, tails, and double nuclei unambiguously signal major mergers which can trigger intense starbursts and central mass growth \citep{Clements1996IRASULIRGsHowManyAreMergers, Bridge2010Mergers, daCunha2010LocalULIRGsPropertiesFromUVtoIR, Ellison2013PostMergers, Tacchella2016HighzCompaction&Quenching}. Stellar bars may facilitate the consumption of fuel for SF and induce the growth of the central bulge, but may be destroyed by major merging \citep{ElicheMoral2018FormationOfS0sAndAssociatedSubstructure, Cavanagh2022BarredGalaxiesEAGLE}. Faded spiral arms in passive disk galaxies may suggest environmental quenching from gas stripping, strangulation, or starvation \citep{Gunn&Gott1972InfallOfMatterIntoClustersAndEffectsOnEvolution, Elmegreen2002ArmStructureAnemicSpirals, Fujita2004PreprocessingGalaxiesBeforeClusterEntry, Masters2010GalaxyZooPassiveRedSpirals}. The properties of the bulge may also suggest different evolutionary pathways. Classical bulges have de Vaucouleurs-like profiles with high S\'ersic indices ($n \gtrsim 2$) whereas pseudo-bulges have lower (i.e., more disky) S\'ersic indices ($n \lesssim 2$) \citep{Drory2007BulgePropsAndGalaxyBimodality, Fisher&Drory2008PseudoVersusClassicalBulgesSersicIndices}. Classical bulges are likely formed by merging whereas pseudo-bulges are likely formed through secular processes \citep{Kormendy2004PseudobulgeReview}. However, to determine the bulge properties requires fitting multiple S\'ersic profiles to a galaxy to decompose it into a bulge and disk \citep[][]{Sersic1963Profile, Freeman1970DisksofSpirals&S0s, Peng2002GALFIT, Simard2011BulgeDiskDecompCatalog, Meert2015SDSSBulgeDiskDecompCatalog}, the robustness of which depends greatly on the resolution and sensitivity of the images. 


To reliably quantify galaxy structure requires images with high resolution and sensitivity in addition to statistically significant sample sizes. However, typical ground-based optical images are limited to resolutions of $\sim1$\arcsec{}. Furthermore, wide-area surveys which are needed for statistical samples are by necessity limited in sensitivity. Small-scale studies have already shown that galaxies which are apparently featureless in SDSS imaging typically have complex and diverse structure in HST imaging \citep{Salim2012SFinOpticallyRedETGs}. The limited resolution and sensitivity of ground-based optical surveys may result in imprecise or biased estimation of structural parameters like the global S\'ersic index as well as the B/T ratio and bulge-specific S\'ersic 
index obtained from bulge-disk decomposition of galaxy profiles \citep{MendelSimard2014BulgeDiskDecompAndStellarMassesForSDSS}. These fundamental properties must be accurately constrained to reliably investigate the morphology-quenching connection and extend the investigation to even finer substructure.


The Hubble Space Telescope (HST) has a resolution of $\sim 0.1$\arcsec{}, an order of magnitude improvement over ground-based optical telescopes.  However, HST's field-of-view is far more modest than SDSS and a dedicated survey of a statistical sample of SDSS galaxies would be cost-prohibitive. Such an endeavor is not needed, however, given that many SDSS galaxies already have `serendipitous' imaging from various smaller HST programs over the years.  We can therefore get the best of both worlds by combining SDSS statistical samples with this serendipitous HST imaging.  

In this work we focus on a sample of $1,743$ SDSS galaxies with robust physical parameters from SED fitting of SDSS and GALEX photometry, S\'ersic profile fits based on SDSS \textit{r}-band imaging, and which also have images from HST.  Using this sample we will investigate the limitations of ground-based optical imaging in quantifying galaxy structure by direct comparison of morphological measurements from both SDSS and HST. Our results will be useful for assessing the limitations of not only ground-based optical surveys but also potentially high-redshift surveys which may be similarly limited in spatial resolution and sensitivity.  

We describe the data and sample selection in Section \ref{Section:Data&SS}.  A full description of our methods is given in Section \ref{Section:Methods}.  We present our results in Sections \ref{Section:HSTAssessment} and \ref{Section:Results} and conclude in Section \ref{Section:Conclusions}.  We adopt a flat WMAP9 cosmology throughout ($H_0 = 69.3$ km/s/Mpc, $\Omega_m = 0.287$) \citep{Hinshaw2013WMAP9Cosmology}.  

\section{Data and Sample Selection}
\label{Section:Data&SS}

In this work we focus on a subset of SDSS (DR10) galaxies taken from the GALEX-SDSS-WISE Legacy Catalog \citep[GSWLC][]{Salim2016GSWLC, Salim2018DustAttCurves} which have optical HST imaging.  GSWLC provides stellar masses and star formation rates (SFRs) for $\sim 700,000$ galaxies with SDSS spectroscopy and GALEX UV coverage.  We start from GSWLC galaxies so that we have access to robust SFRs and stellar masses which we can use to assess the differences between HST and SDSS in the context of various scaling relations. GSWLC obtained these properties via SED fitting to the broadband UV-optical photometry with added constraints from the WISE-derived IR luminosity whenever available. We use the GSWLC-X2 catalog which is based on the deepest available UV imaging for each object.  To be included in the GSWLC catalog, galaxies were required to have a redshift in the range $0.01 < z < 0.3$ and an \textit{r}-band Petrosian magnitude ($r$) in the range $r < 18$, amounting to $\sim 91\%$ of SDSS DR10 spectroscopic targets. The GSWLC-X2 catalog includes 659,229 total galaxies.  

We make use of SDSS DR16 \citep{Ahumada2020SDSSDR16datarelease} model fits; an exponential disk and a de Vaucouleurs profile. For each object we take the ellipticity and position angle associated with the best-fitting model in the \textit{r} band, which we take to be the one with a higher likelihood value. The SDSS-derived ellipticity and position angle will be used to inform the initial guess required for the HST profile fitting. 

We also make use of the \citet[][hereafter S11]{Simard2011BulgeDiskDecompCatalog} catalog of galaxy structural properties. \citetalias{Simard2011BulgeDiskDecompCatalog} includes data for $\sim 1$ million SDSS galaxies in DR7. From SDSS images \citetalias{Simard2011BulgeDiskDecompCatalog} obtained single S\'ersic fits in addition to two different bulge-disk decompositions, one where the S\'ersic index of the bulge was fixed at $n = 4$ and another where the bulge S\'ersic index was allowed to vary freely. In both cases the disk was modeled as exponential ($n=1$). $\sim 96\%$ of GSWLC galaxies are included in \citetalias{Simard2011BulgeDiskDecompCatalog}. We adopt model parameters derived from the \textit{r}-band imaging to use as a basis of comparison for HST-derived parameters. 

We additionally compare to the catalog of \citet[][hereafter M15]{Meert2015SDSSBulgeDiskDecompCatalog}. \citetalias{Meert2015SDSSBulgeDiskDecompCatalog} uses a similar fitting procedure to \citetalias{Simard2011BulgeDiskDecompCatalog}, performing single S\'ersic as well as two-component bulge/disk decompositions with both fixed ($n=4$) and free ($n$ varies) bulges plus exponential disks for $\sim 700,000$ galaxies in SDSS DR7. In contrast to \citetalias{Simard2011BulgeDiskDecompCatalog}, \citetalias{Meert2015SDSSBulgeDiskDecompCatalog} uses a different fitting code ({\tt GALFIT} instead of {\tt GIM2D}) and makes some methodological improvements over \citetalias{Simard2011BulgeDiskDecompCatalog} such as a more robust sky value estimation. $\sim 98\%$ of GSWLC galaxies are included in \citetalias{Meert2015SDSSBulgeDiskDecompCatalog}. As with \citetalias{Simard2011BulgeDiskDecompCatalog}, we adopt \citetalias{Meert2015SDSSBulgeDiskDecompCatalog} model parameters derived only from the \textit{r}-band imaging. 

Finally, we also compare to properties from the catalog of \citet[][hereafter NYU]{Blanton2005NYUcatalogForSDSS}. The NYU catalog includes single-component axisymmetric S\'ersic fits for $\sim2.5$ million galaxies from various surveys including SDSS DR2, which we use as another point of comparison for the HST-derived single-component profiles. $\sim 61\%$ of GSWLC galaxies are included in the NYU catalog; a large fraction of GSWLC galaxies are unmatched because NYU only includes SDSS DR2. As with \citetalias{Simard2011BulgeDiskDecompCatalog} and \citetalias{Meert2015SDSSBulgeDiskDecompCatalog}, we adopt NYU model parameters derived from \textit{r}-band imaging. 




\subsection{HST}

\begin{figure}
    \centering
    \includegraphics[scale=0.3]{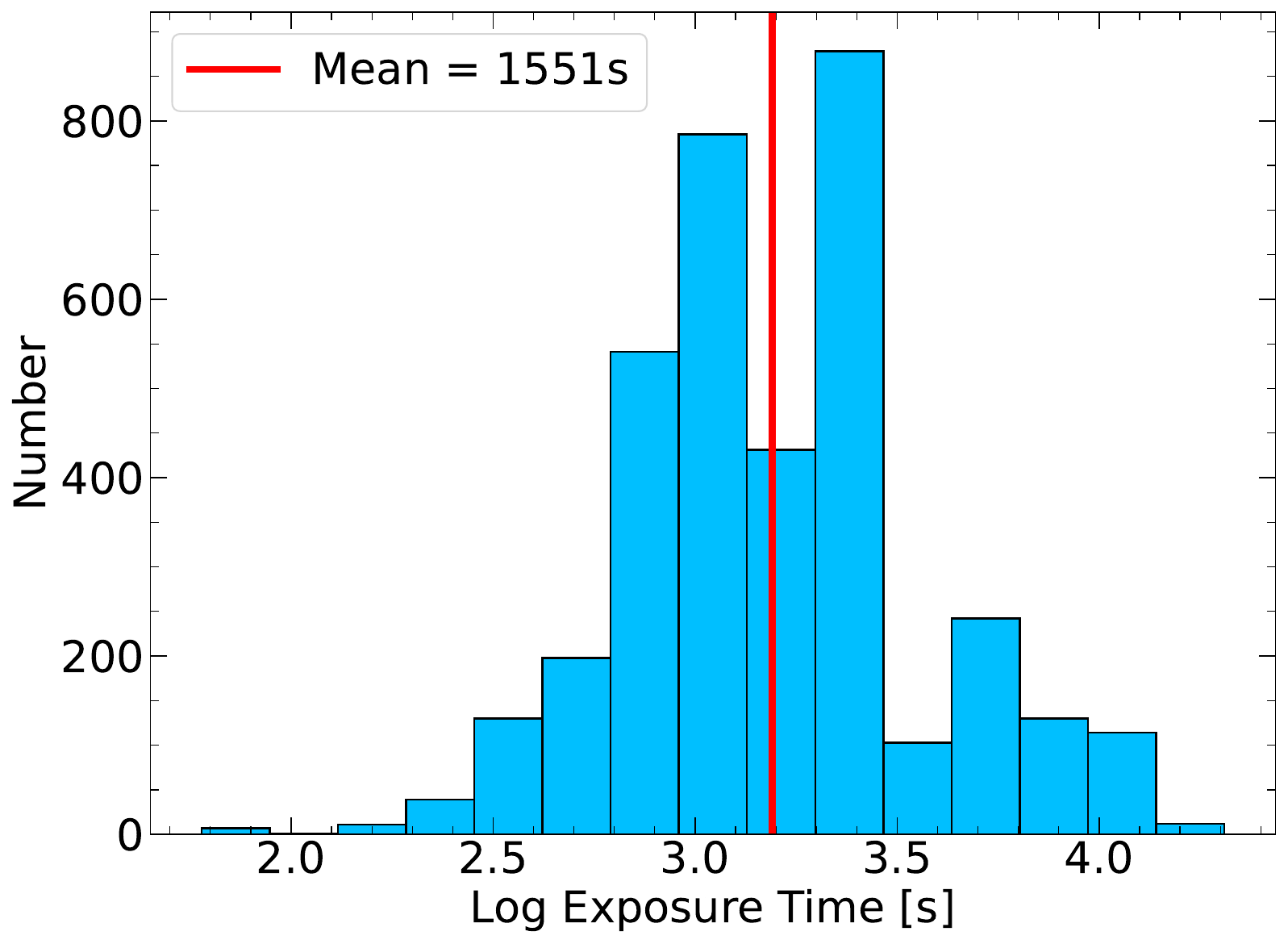}
    \caption{Distribution of exposure times (in log scale) for the sample of SDSS (GSWLC) galaxies with HST imaging considered in this work. The mean exposure time is 1,551 seconds and is shown as a red line.}
    \label{GSW_exposure_time_dist}
\end{figure}

\begin{figure}
    \centering
    \includegraphics[scale=0.3]{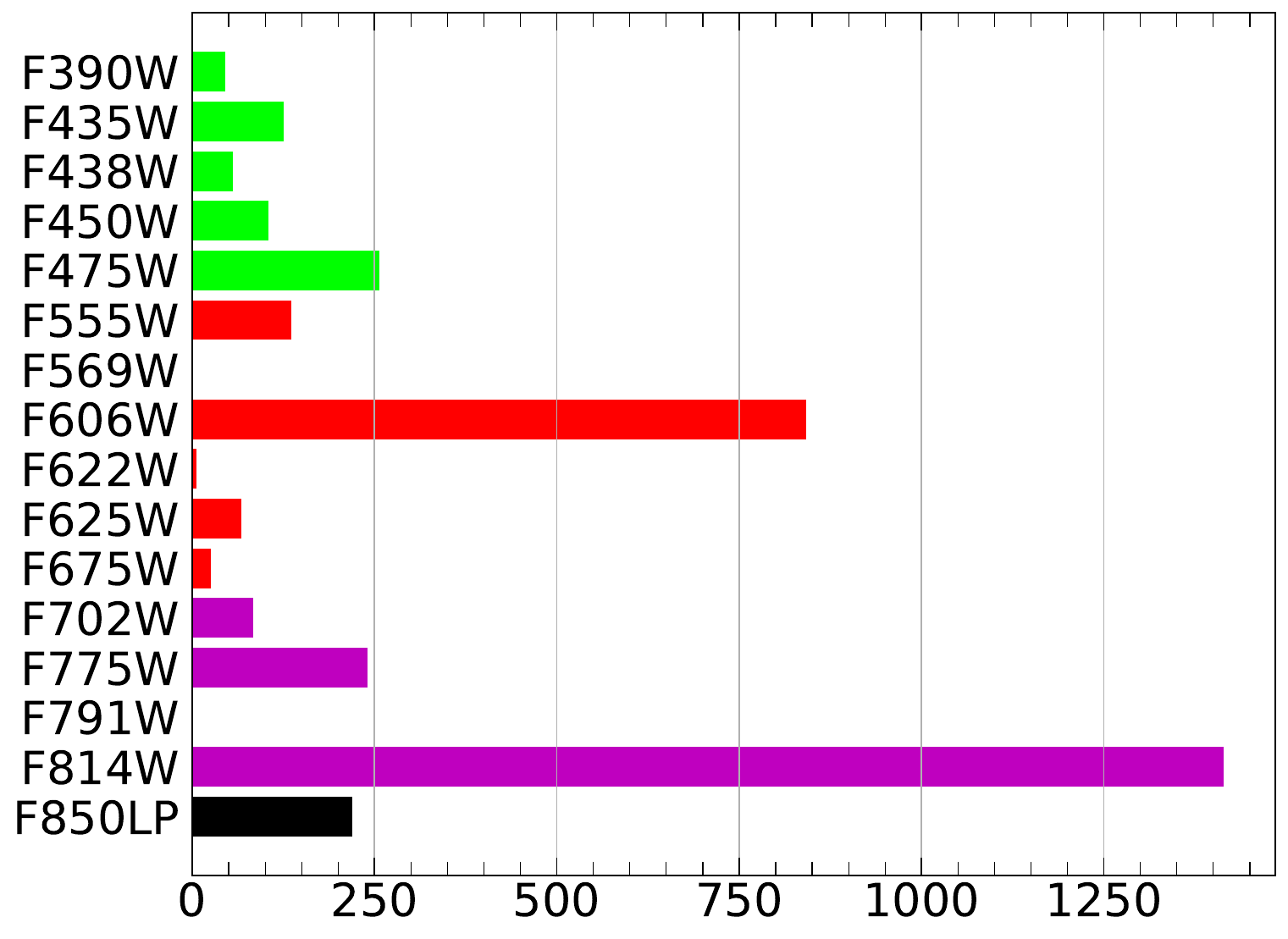}
    \caption{Distribution of filters for the sample of SDSS (GSWLC) galaxies with HST imaging considered in this work.  Filters are colored according to the closest SDSS filter in terms of wavelength; green, red, magenta, and black represent \textit{g}, \textit{r}, \textit{i}, and \textit{z} respectively.}
    \label{GSW_filters_dist}
\end{figure}

\begin{figure}
    \centering
    \includegraphics[scale=0.3]{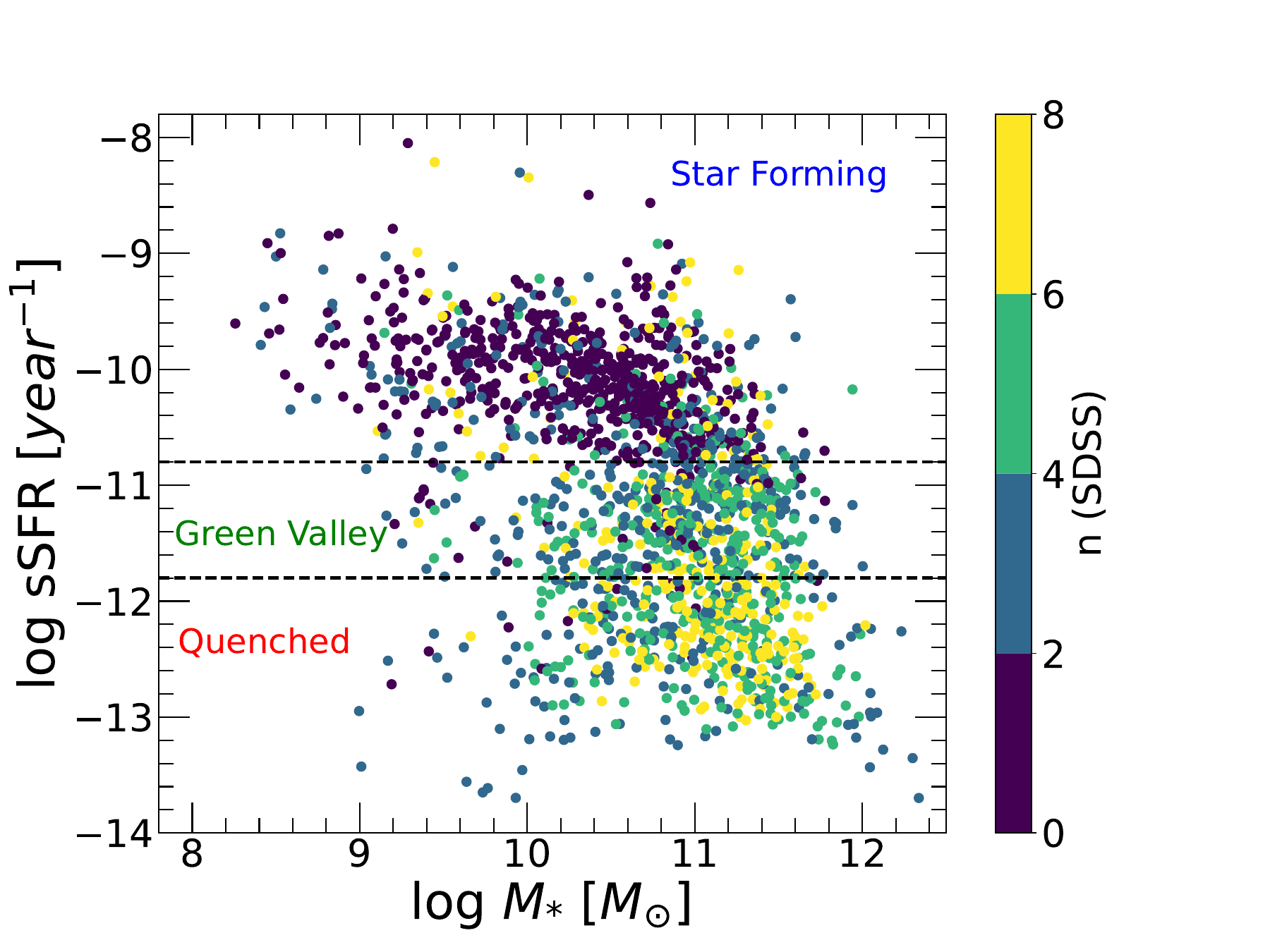}
    \caption{Distribution in specific SFR (sSFR) versus stellar mass ($M_*$) for the sample of SDSS (GSWLC) galaxies with HST imaging considered in this work. Points are colored by the SDSS-derived global S\'ersic index from \citetalias{Simard2011BulgeDiskDecompCatalog}. The dashed lines separate star-forming, green valley, and quenched galaxies following \citet{Salim2014GreenValley}.}
    \label{GSW_sSFRvM_ncolor_dist}
\end{figure}

We obtain imaging data from the Hubble Legacy Archive (HLA) and use source lists from the Hubble Source Catalog (HSC, version 3) \citep{Whitmore2016HubbleSourceCatalog} for cross-matching source positions from SDSS to identify HST images of SDSS objects. 

Using MAST Casjobs, we select HST sources within 3\arcsec{} of any SDSS position which have images in HST optical wide bands from F390W to F850LP.  Specifically, we limit our consideration to the following set of filters: F390W, F435W, F438W, F450W, F475W, F555W, F569W, F606W, F622W, F625W, F675W, F702W, F775W, F791W, F814W, and F850LP.  Having a span of filters will allow us to assess the dependence of morphological properties on wavelength (i.e., the k-correction) for galaxies with images in multiple bands. The imaging comes from a variety of instruments: WFPC2, ACS, and WFC3. Their FWHM ranges from 0.1\arcsec{} to 0.2\arcsec{}, compared to 1.4\arcsec{} for SDSS. In total, 1,988 GSWLC objects have at least one HST image in our selection of filters.  


Using the image names from the SDSS and HSC cross-match, we download image cutouts from HLA using the {\tt fitscut} interface.  Images are centered on the RA/DEC position reported by SDSS, and the cutout size is set at six times the SDSS Petrosian radius, which ensures a significant number of background pixels and avoids excluding faint outer regions of the galaxy which SDSS may not detect.  Images with $> 5\%$ of bad (i.e., `nan' valued) pixels within a square sub-cutout of width equal to twice the SDSS Petrosian radius centered on the SDSS position are excluded from our analysis; this avoids most galaxies cut off by image bounds or chip gaps.  Applying the bad pixel cut leaves us with 1,743 galaxies with reliable HST optical imaging.  



Figure \ref{GSW_exposure_time_dist} shows the distribution of exposure times for all HST images associated with our sample.  The mean exposure time is 1,551 seconds, while the shortest exposure time is $\sim 100$ seconds.  We note that the exposure time for SDSS images is 53.9 seconds, so the images in our sample typically represent a significant improvement in depth over the SDSS imaging.  Figure \ref{GSW_filters_dist} shows the distribution of filters for all HST images associated with our sample, colored roughly by the closest SDSS filter in terms of wavelength.  The most common filters are F606W and F814W by a wide margin. 

Figure \ref{GSW_sSFRvM_ncolor_dist} shows the log specific star formation rate (sSFR) versus the log stellar mass ($M_*$), both taken from GSWLC, for the galaxies in our sample. Galaxies are also colored by the global S\'ersic index from \citetalias{Simard2011BulgeDiskDecompCatalog}. The sSFR thresholds separating star-forming, green valley, and quenched galaxies are shown as dashed lines, and adopted from \citet{Salim2014GreenValley}. Our sample spans a wide range in mass, sSFR, star formation activity, and morphology. 



\section{Methods}
\label{Section:Methods}

\begin{figure}
    \centering
    \includegraphics[scale=0.315]{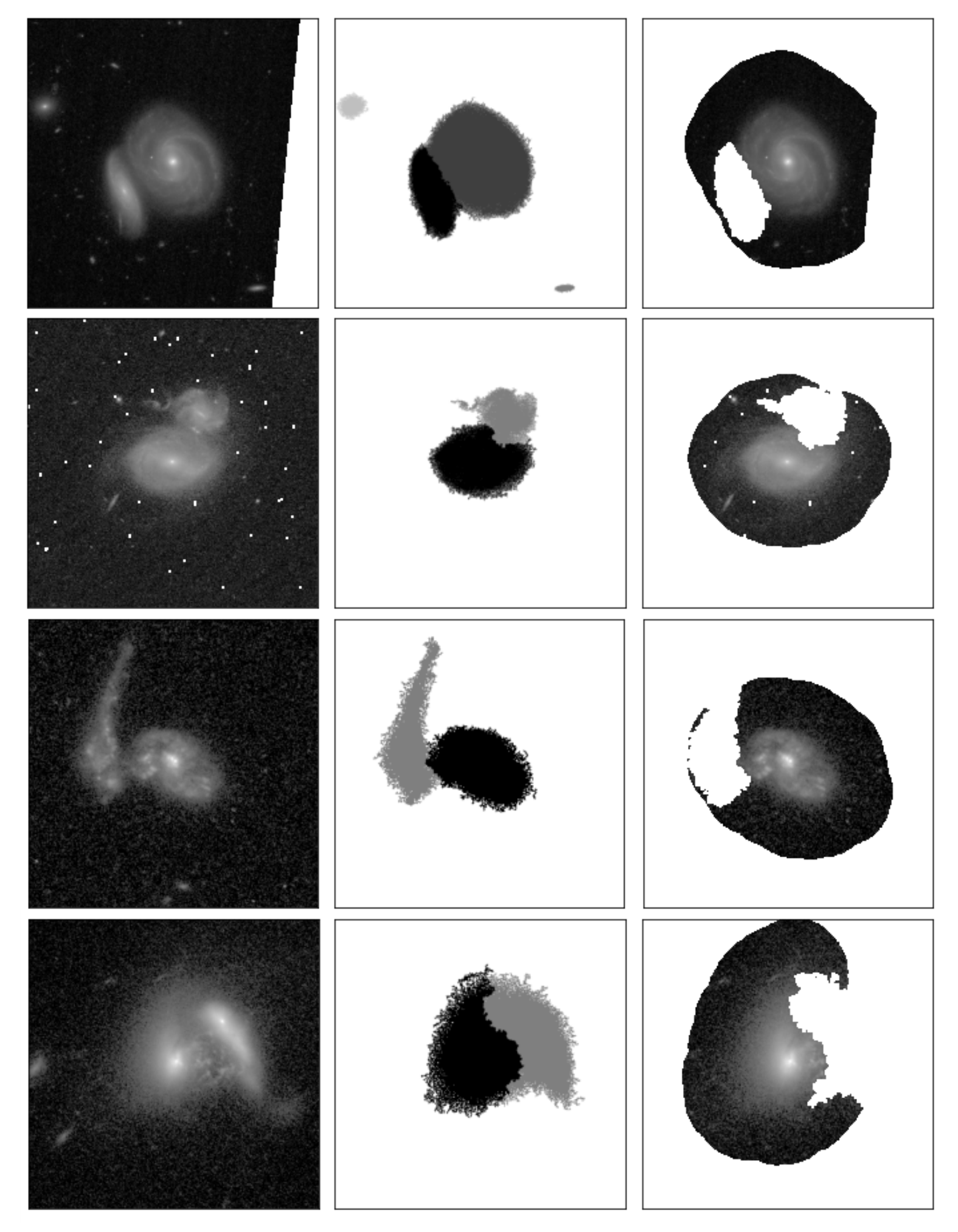}
    \caption{Example images showing our deblending and masking procedure. The background-subtracted image is shown on the left, the deblended segmentation map in the middle, and the final masked image on the right. Bright neighboring object pixels and distant ($> 1$ SDSS petrosian radius) sky pixels are masked from the fitting. All models are fit to the entire masked image (object and sky) to include faint object pixels that may fall under the threshold used for the segmentation.}
    \label{example_improcess}
\end{figure}

\subsection{Image processing}


A segmentation map, which assigns pixels to either the sky or to individual objects in the image, is needed to mask out bright neighbors and estimate robust sky background levels. We use Astropy's {\tt photutils} package to perform the image segmentation. The segmentation algorithm requires an estimate of the mean sky value and a sigma value (i.e., some multiple of the sky standard deviation) as input. Source detection is performed only for pixels with values above the mean plus the sigma value. We use $3\sigma$ clipping on the entire image to estimate the mean and standard deviation of the sky pixels. The sigma value in the segmentation algorithm is taken to be the sky standard deviation ($1\sigma$ threshold). We set the minimum number of pixels for each object to be 1000, and set the connectivity value to 4 (so object pixels must be touching on at least one side to be connected).  Deblending is applied to separate objects with overlapping segmentation maps, for which we set the number of thresholding levels at 32 and set the contrast parameter to 0.01 in addition to keeping the same requirement of at least 1000 pixels per object. We note that the profile fitting is not limited to object pixels in the segmentation map, since the segmentation map may not include faint outer regions of the target galaxy. 


To perform the background subtraction, we obtain a more robust estimate of the mean sky value by performing $1\sigma$ clipping on the subset of image pixels identified as sky by the segmentation map.  We verify the accuracy of the sky background from this method by visual inspection of the image pixel histograms for $\sim 100$ different images and find that the means obtained using this method are highly robust at capturing the observed peaks in the image histograms.  We subtract the mean sky value from the image to obtain the background-subtracted image used for the profile fitting.

We also require a noise or `sigma' ($\sigma$) map to be used in the profile fitting. The noise map is used to calculate the weight of each pixel in the fitting (i.e., the weight map). To obtain a noise map we first estimate the object noise map as the square root of the background-subtracted image, with negative pixel values in the image set to zero (so that the object flux or object noise for the negative pixel is zero). We then add to the object noise map an estimate of the sky noise in quadrature, assuming that the sky noise is the same for every pixel.  We assume the sky noise to be equal to the sky mean used in the background subtraction, if greater than zero. For a very small number of images ($1\%$) the sky mean was determined to be less than zero, likely due to errors in the bias subtraction. To get an estimate of the sky noise in such cases, we perform $5\sigma$ clipping on the entire image and adopt the standard deviation obtained from the clipping as the sky noise instead.

We mask out all object pixels which do not belong to the target (as determined by the segmentation map) to avoid the inclusion of bright regions of neighboring galaxies in the profile fitting.  We also mask any pixels farther than one SDSS petrosian radius (i.e., 1/6th of the image size) away from an object pixel belonging to the target to minimize the contamination from image artifacts and/or faint regions of neighboring galaxies which may lie just beyond the bounds of the image.  We also mask all sky pixels which lie close to an image boundary or chip gap (specifically, within $10\%$ of the image size) as such regions commonly contain bright artifacts. The fitting is performed on the entire masked image. We show examples of background-subtracted images, segmentation maps, and masked images in Figure \ref{example_improcess}. 

\begin{figure*}
    \centering
    \gridline{\fig{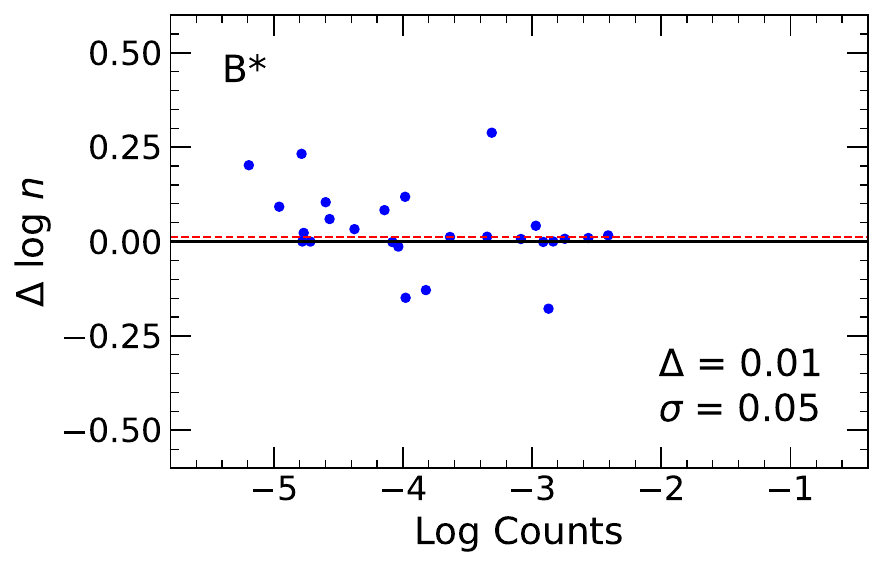}{0.3\textwidth}{}
          \fig{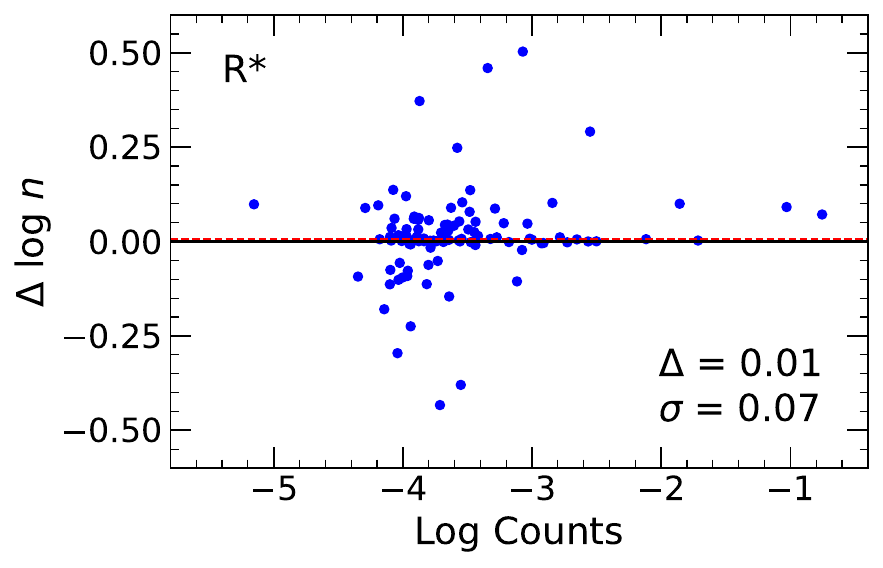}{0.3\textwidth}{}
          \fig{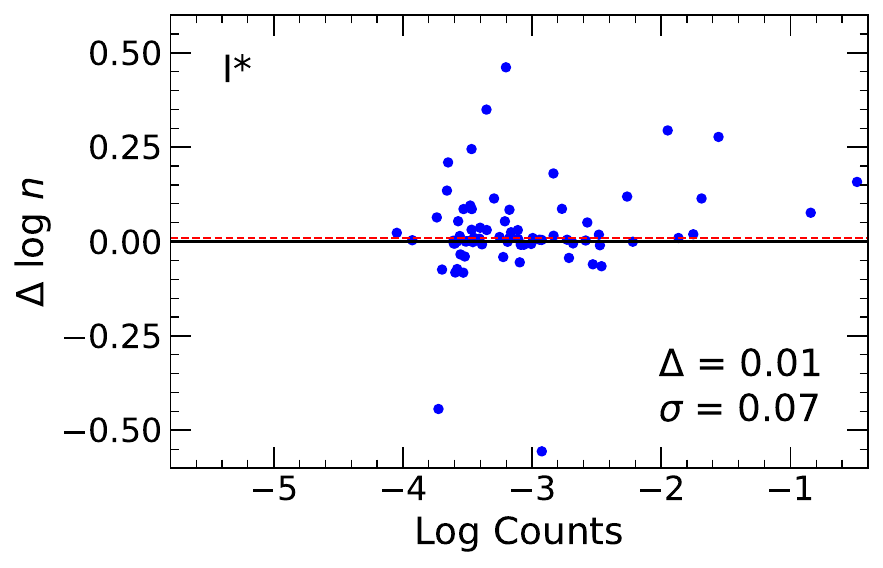}{0.3\textwidth}{}
          }
    \vspace*{-\baselineskip}
    \gridline{\fig{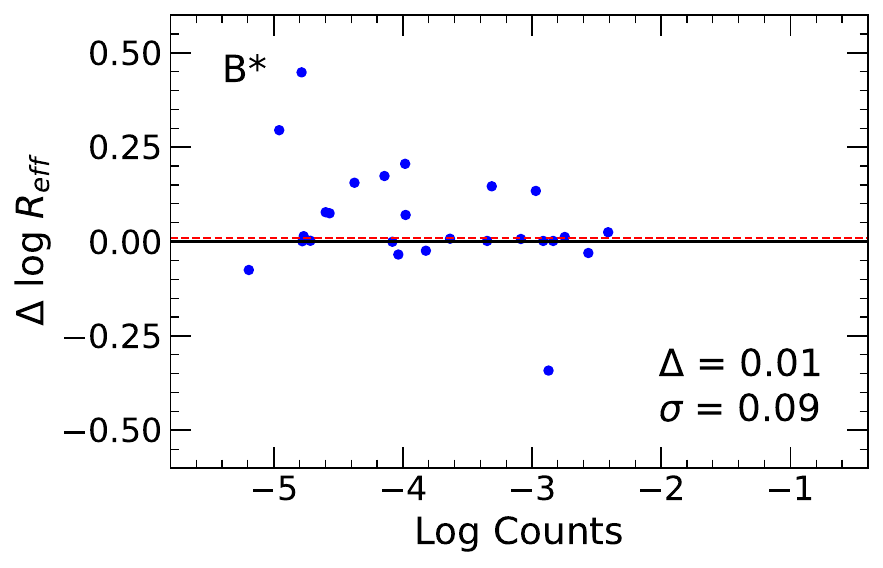}{0.3\textwidth}{}
          \fig{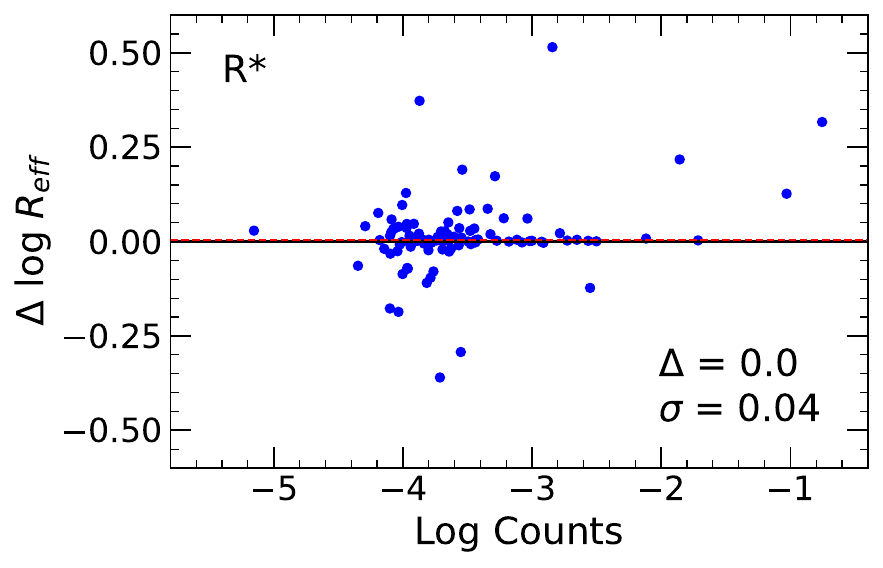}{0.3\textwidth}{}
          \fig{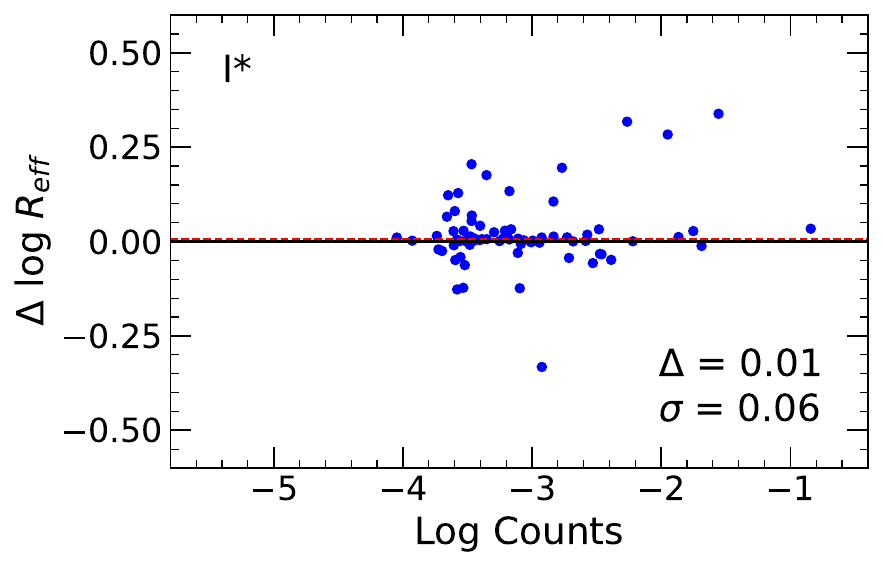}{0.3\textwidth}{}
          }
    \vspace*{-\baselineskip}
    \gridline{\fig{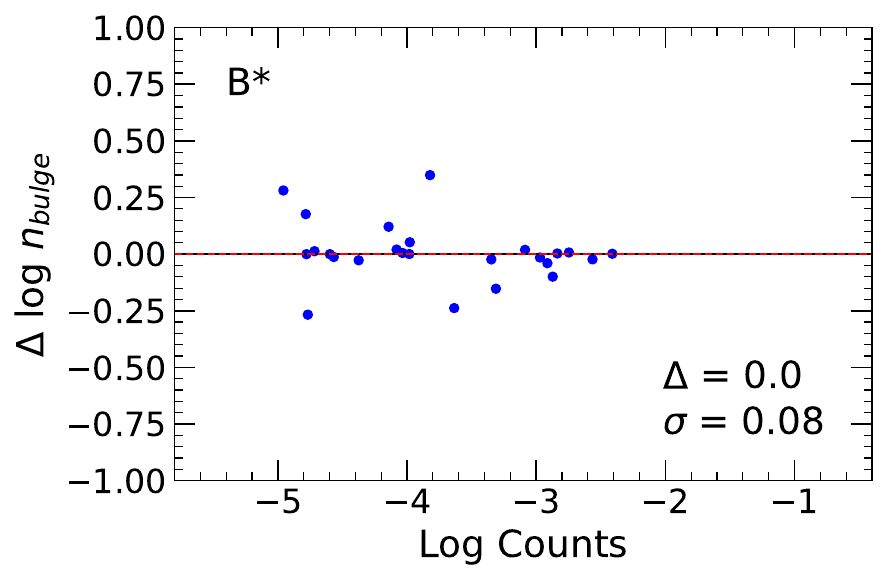}{0.3\textwidth}{}
          \fig{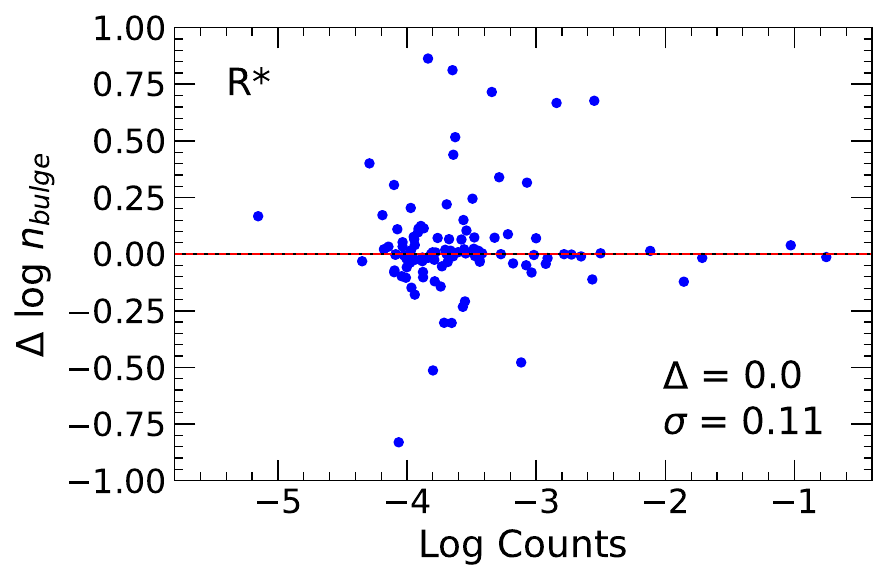}{0.3\textwidth}{}
          \fig{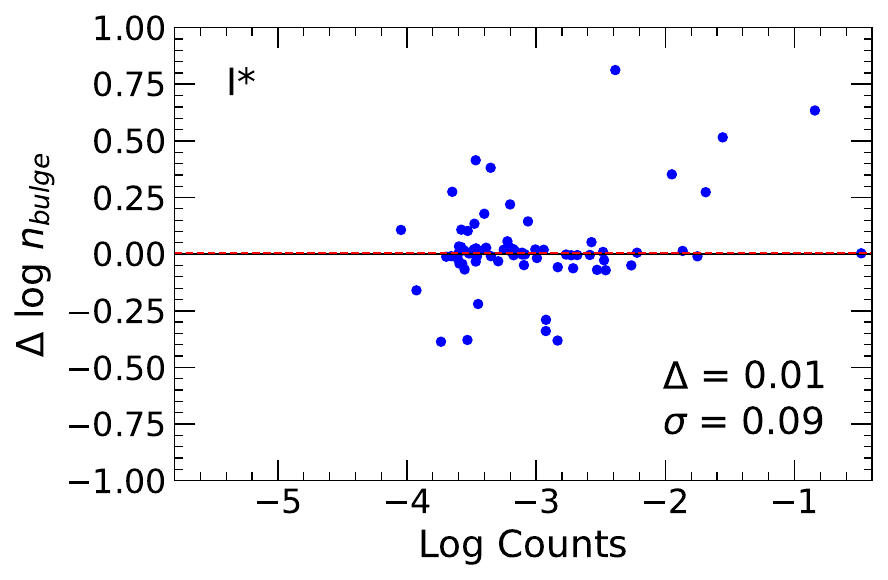}{0.3\textwidth}{}
          }
    \vspace*{-\baselineskip}
    \gridline{\fig{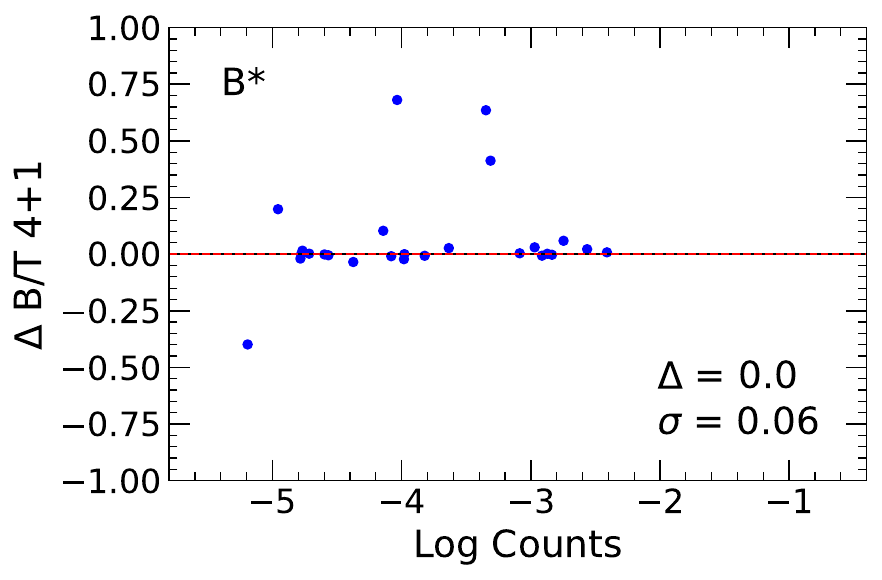}{0.3\textwidth}{}
          \fig{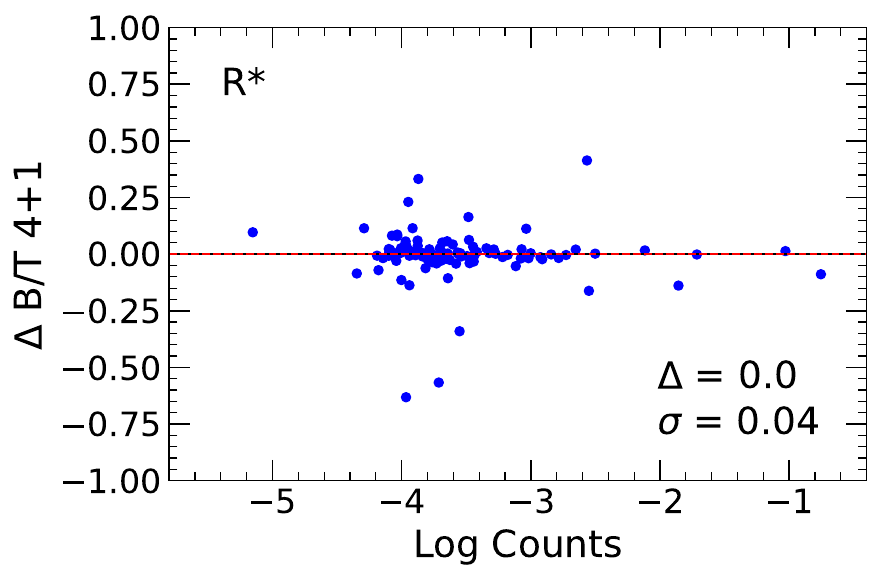}{0.3\textwidth}{}
          \fig{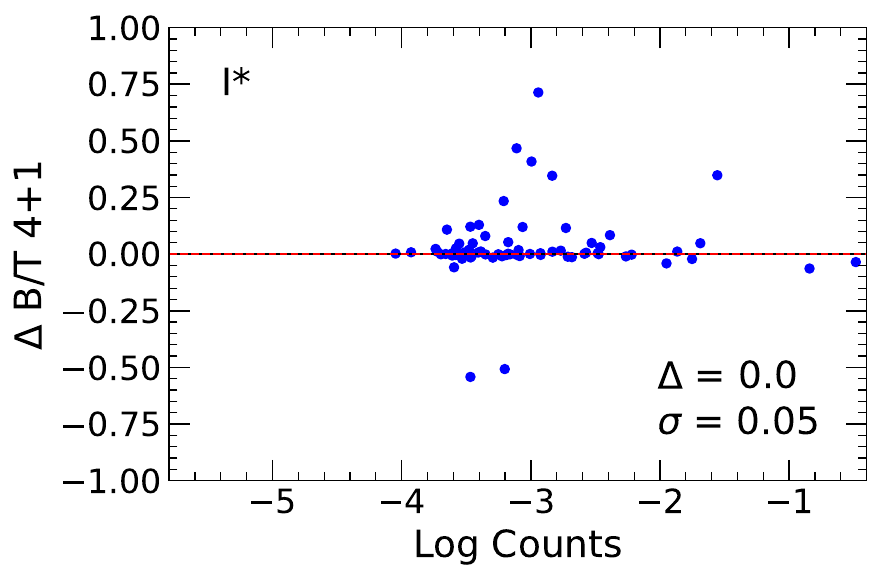}{0.3\textwidth}{}
          }
    \vspace*{-\baselineskip}
    \gridline{\fig{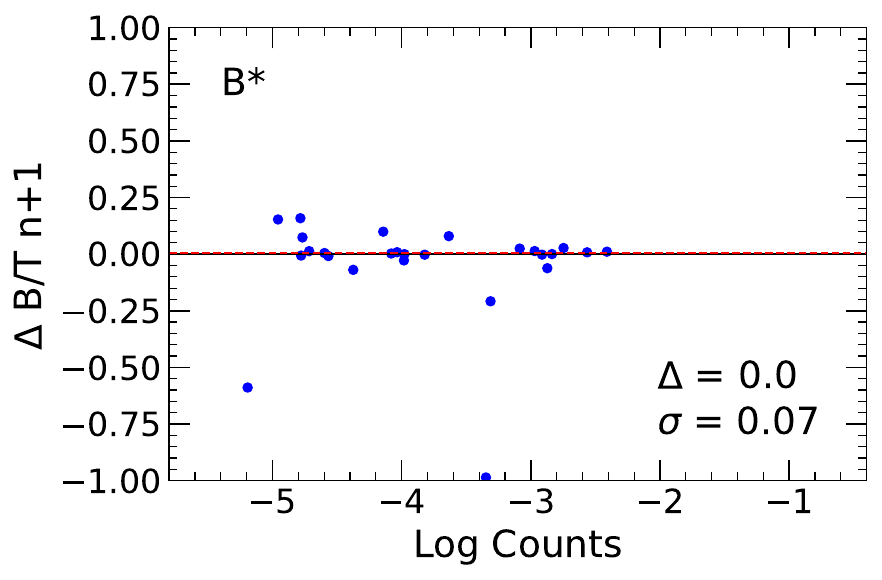}{0.3\textwidth}{}
          \fig{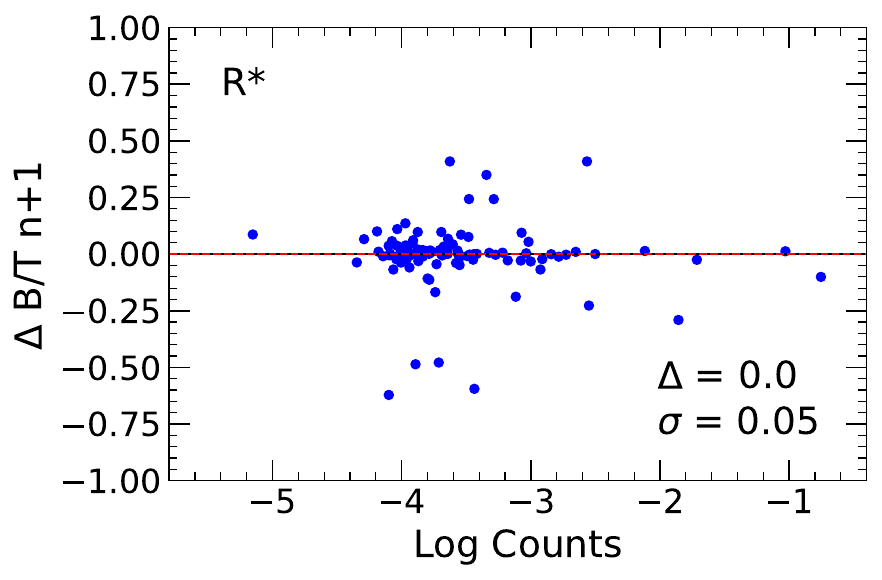}{0.3\textwidth}{}
          \fig{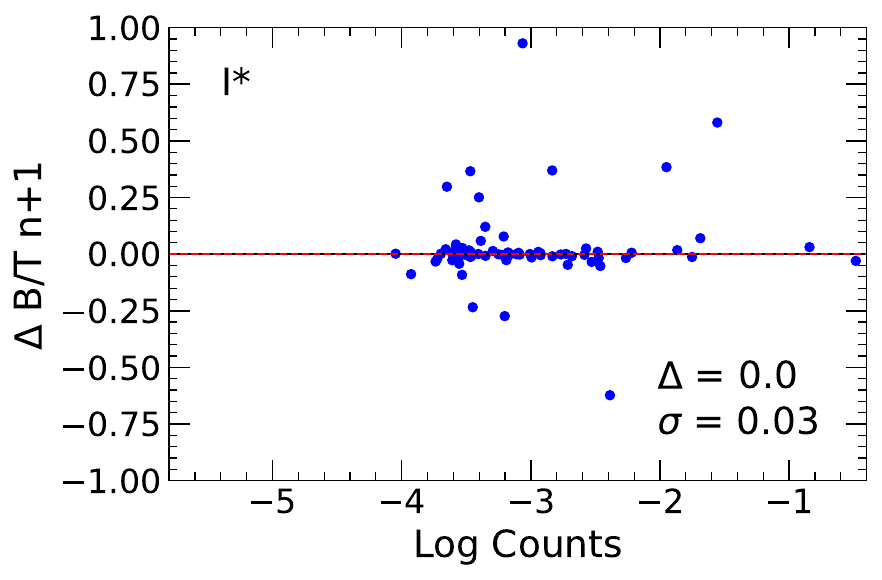}{0.3\textwidth}{}
          }
    \vspace*{-\baselineskip}
    \caption{Difference in HST-derived morphology metrics (log global $n$, log half-light radius ($R_{\text{eff}}$), log bulge $n$, B/T (4+1), B/T (n+1)) between image pairs for galaxies with at least two images at a similar wavelength versus the single-component model flux (in total photon counts). Differences are between the two deepest images in a given band group ($B^*$ left, $R^*$ middle, and $I^*$ right) and is the property of the fit with greater flux minus that of the fit with smaller flux. The flux on the x-axis is the smaller of the two values. Medians and robust standard deviations (from 16th and 84th percentiles) are given as $\Delta$ and $\sigma$ respectively.}
    \label{BRI_self_consistency_plots}
\end{figure*}



\subsection{S\'ersic profile fitting and bulge/disk decomposition}

We fit three different types of profiles to each galaxy in a similar manner to \citetalias{Simard2011BulgeDiskDecompCatalog}. We use 2D S\'ersic profiles defined by a S\'ersic index ($n$), effective radius ($R_{\text{eff}}$, profile amplitude, profile centroid, position angle, and ellipticity. The first is a single-component (global) S\'ersic profile. The second is a two-component S\'ersic profile where the bulge S\'ersic index ($n_{\text{bulge}}$) is fixed at $n = 4$ and the disk S\'ersic index is fixed at $n = 1$. The third is a two-component S\'ersic profile where the bulge S\'ersic index is allowed to vary while the disk S\'ersic index is still fixed at $n = 1$. Herafter, we refer to the profile for which $n_{\text{bulge}}=4$ as the `4+1' fit and refer to the profile where $n_{\text{bulge}}$ varies as the `n+1' fit. 

We use the {\tt Petrofit} package \citep{PETROFIT_Geda_2022} to perform the profile modeling and fitting.  {\tt Petrofit} is a Python wrapper that includes many tools useful for galaxy photometry and profile fitting.  For this work we use the {\tt Petrofit} S\'ersic model fitting functions which are capable of fitting an arbitrary number of S\'ersic components (e.g., disks and bulges). The fit is performed on the masked image, for which pixels assigned to neighboring objects by the segmentation map and sky pixels close to image boundaries are masked out to avoid biasing the fits. {\tt Petrofit} assumes that the weight map is the reciprocal of the noise map (so weights are $1 / \sigma$). Optionally a PSF may be included; we compute PSFs for the HST images using TinyTim \citep{Krist2011TinyTimHSTPSFs} and input them to {\tt Petrofit}. 

{\tt Petrofit} requires an initial guess for each model parameter. We use the Python package {\tt statmorph} \citep{Statmorph2019MNRAS.483.4140R} to obtain Petrosian shapes and photometry from the HST images which we use to inform the initial guesses for {\tt Petrofit}. We note that unlike {\tt Petrofit}, {\tt statmorph} adopts the noise map as the weight map (so for each pixel the weights are simply $\sigma$). 

For the single component fits with {\tt Petrofit} we take the initial half-light radius to be the elliptical half-light Petrosian radius determined by {\tt statmorph}. The initial global S\'ersic index ($n$) is defined based on the concentration index (\textit{C}) as measured by {\tt statmorph} using the following empirical relation: $n = 10^{-1.5} \times C^{3.5}$. This relation was taken from {\tt statmorph}, which uses it internally for its single-S\'ersic fitting function \citep[not used in this work; see][]{Statmorph2019MNRAS.483.4140R}. We limit the initial S\'ersic index to the range [1, 3.5] to avoid initial guesses close to the extreme ends of the parameter space. The initial ellipticity and position angle are always taken from the best-fitting SDSS pipeline model (see Section \ref{Section:Data&SS}). We take the initial profile centroid (in x and y) to be the value computed from the HST image by {\tt statmorph}. 

For the two-component fits we take the elliptical half-light Petrosian radius as the initial disk radius. The initial bulge radius is taken to be half that of the disk. The initial bulge S\'ersic index ($n_{\text{bulge}}$) is taken to be the global S\'ersic index of the single component fit, but limited to the range [1, 3.5]. The initial bulge-to-total ratio is determined empirically from the concentration index (\textit{C}) using the relation from \citet{Conselice2003CAS} and is limited to the range [0,1]. The initial centroid, ellipticity, and position angle for both components are the same as the initial values for the single-component fit.

For both the single-component and two-component profiles we restrict the fitting range of the S\'ersic index to $0.25 < n < 10$.  Component radii are not allowed to be smaller than 1 pixel but are otherwise unbounded.  Component amplitudes are not allowed to be negative, but are otherwise unbounded.  Ellipticities are allowed to vary between 0 and 1, and position angles are allowed to vary in the range $\pi < \theta < \pi$.  

We allow a degree of `slack' in the profile centroids equal to 10 pixels in x and y. We also find that the reported SDSS positions and the HST-derived Petrosian centroids (from {\tt statmorph}) are systematically offset by $\sim 2$ pixels on average, so we set the initial x and y centroid to the center x and y of the image subtracted by 2 pixels (in both x and y) to account for this offset.

We make a slight modification to the {\tt Petrofit} `fit\textunderscore{}model' function.  By default, {\tt Petrofit} makes use of the Levenberg-Marquardt algorithm to perform the profile fitting.  We found in testing that the Levenberg-Marquardt algorithm can be highly sensitive to small changes in the initial guess. On top of this, the Levenberg-Marquardt algorithm produces unreliable fits for as much as $20\%$ of the galaxies in our sample, with the fits tending to get `stuck' on the boundaries of the parameter space and also sometimes leading to inconsistent profiles for galaxies with multiple images even in cases where the images are similar in passband and exposure time.  These issues may occur due to the very high resolution of our images; since these are relatively nearby galaxies, the typical number of pixels is quite large compared to e.g., SDSS images of similar angular extent.  We therefore modify the {\tt Petrofit} `fit\textunderscore{}model' to instead use Astropy's Trust Region Reflective algorithm (TRFLSQFitter) instead of the Levenberg-Marquardt fitter (LMLSQFitter or LevMarLSQFitter).  We find that the Trust Region algorithm prevents fits from `sticking' to parameter bounds, is relatively insensitive to initial guesses, and produces more consistent fits for multiple images of the same galaxy.  


\begin{figure*}
    \centering
    \gridline{\fig{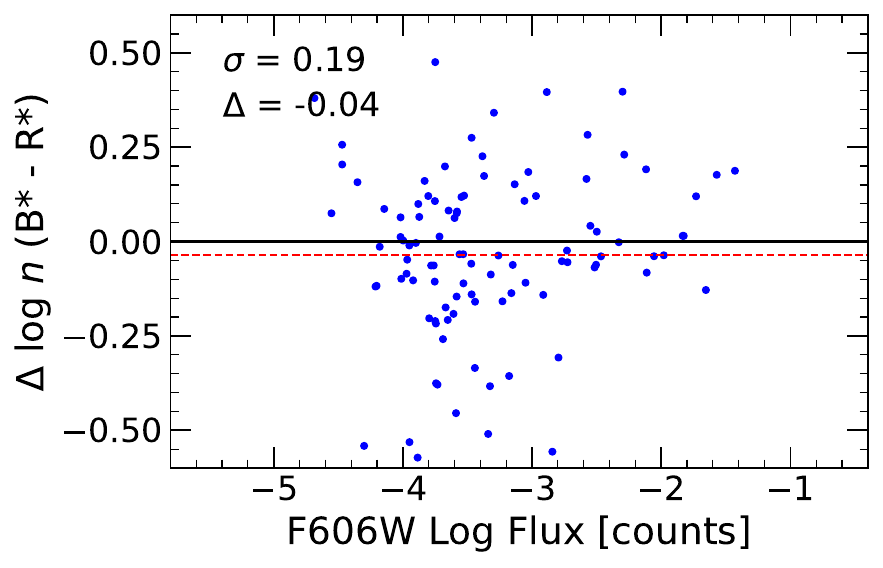}{0.3\textwidth}{}
          \fig{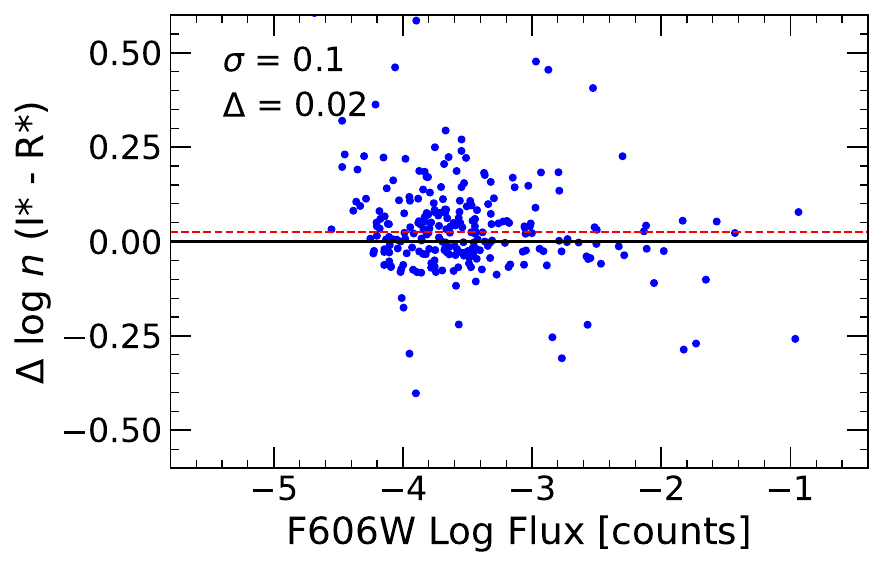}{0.3\textwidth}{}
          \fig{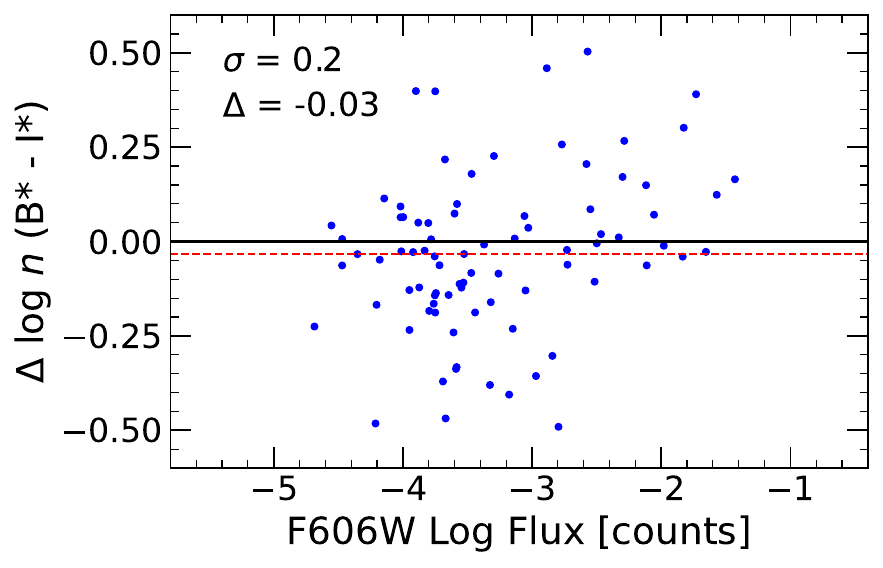}{0.3\textwidth}{}
          }
    \vspace*{-\baselineskip}
    \gridline{\fig{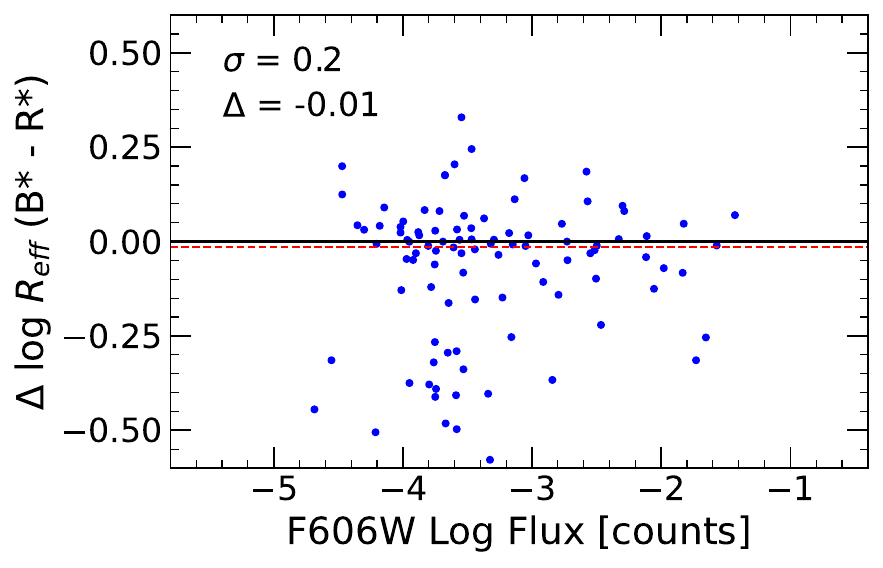}{0.3\textwidth}{}
          \fig{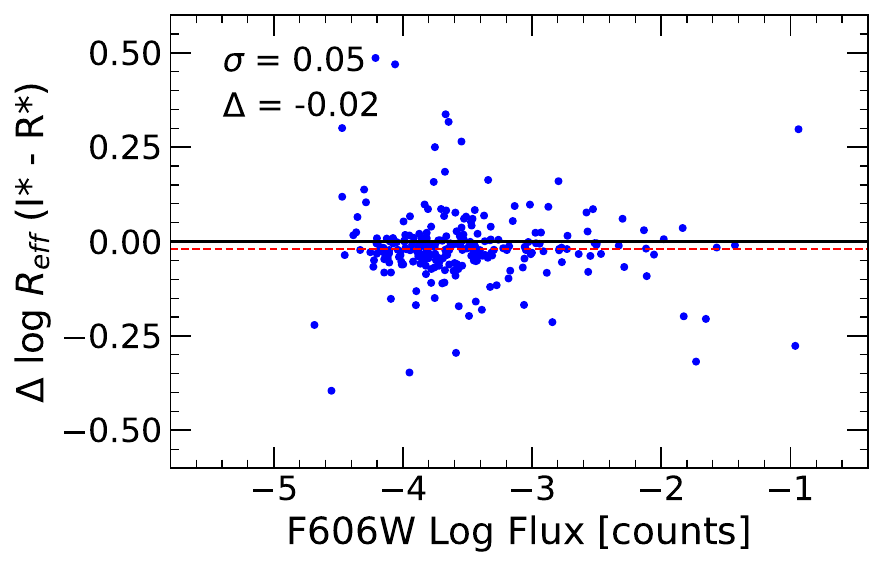}{0.3\textwidth}{}
          \fig{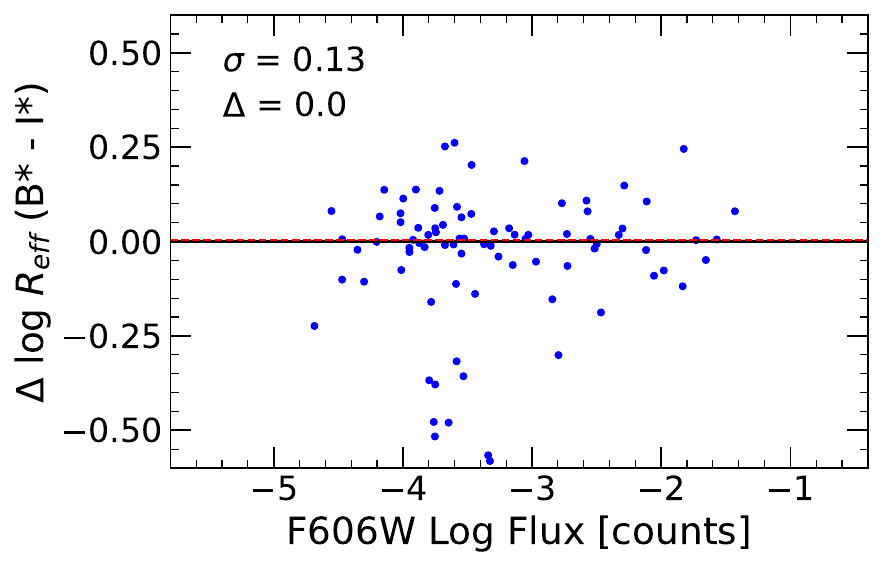}{0.3\textwidth}{}
          }
    \vspace*{-\baselineskip}
    \gridline{\fig{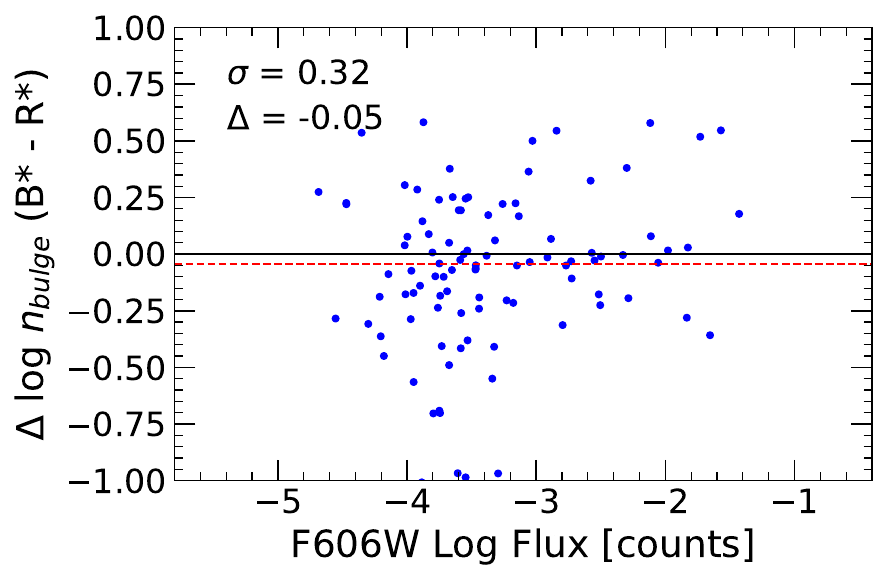}{0.3\textwidth}{}
          \fig{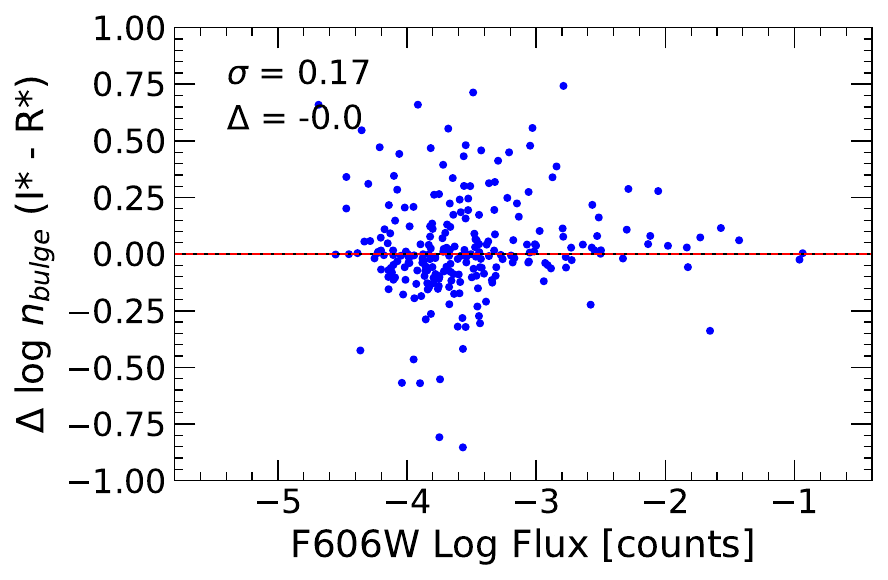}{0.3\textwidth}{}
          \fig{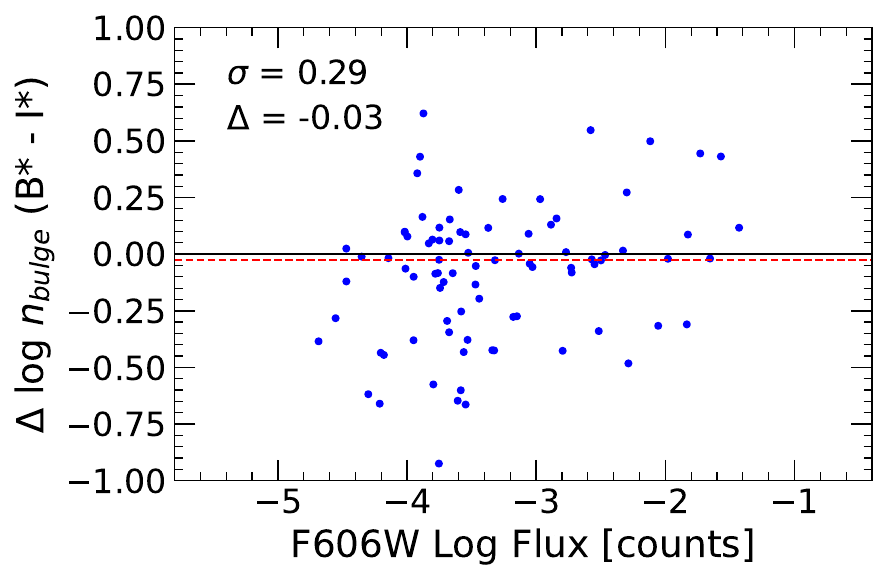}{0.3\textwidth}{}
          }
    \vspace*{-\baselineskip}
    \gridline{\fig{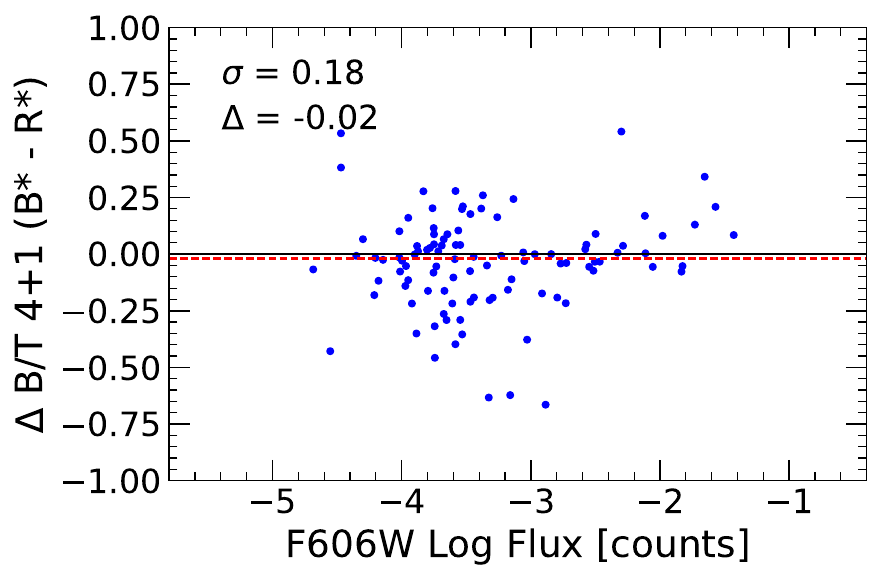}{0.3\textwidth}{}
          \fig{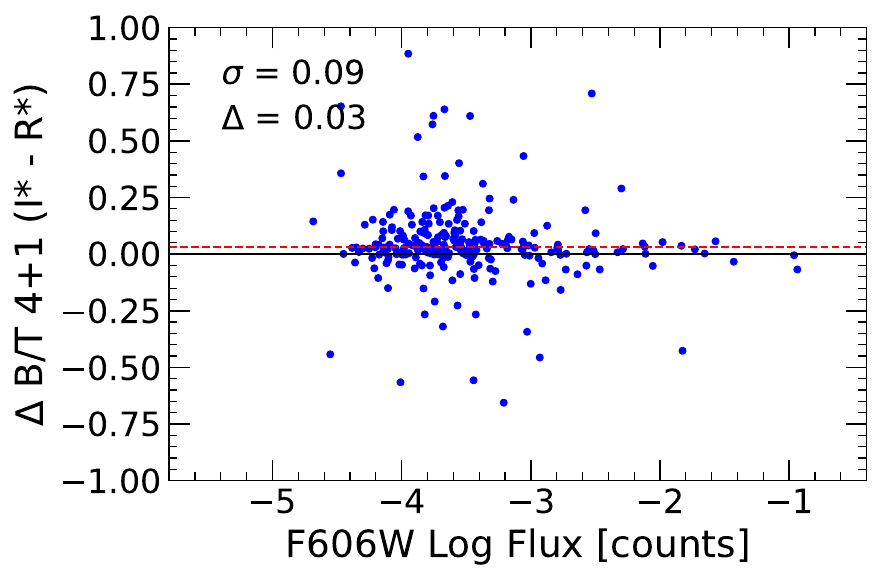}{0.3\textwidth}{}
          \fig{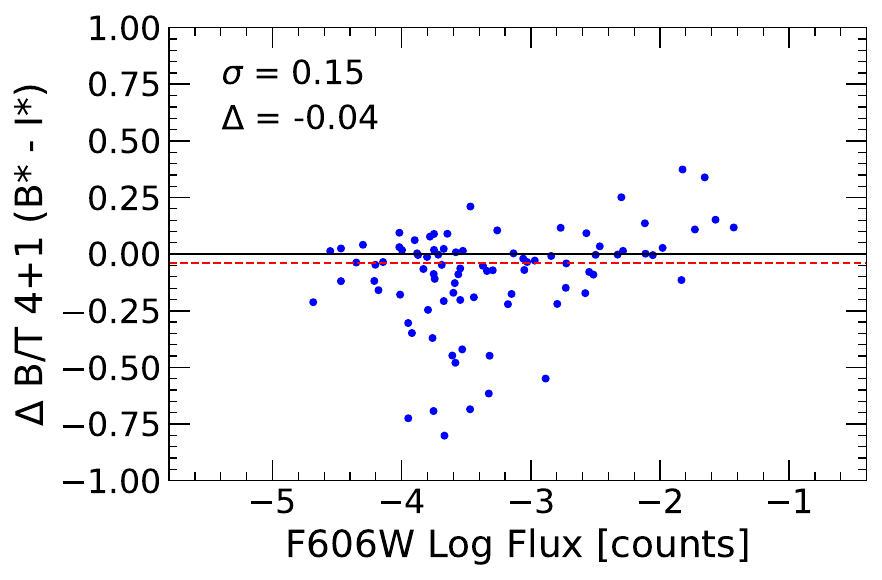}{0.3\textwidth}{}
          }
    \vspace*{-\baselineskip}
    \gridline{\fig{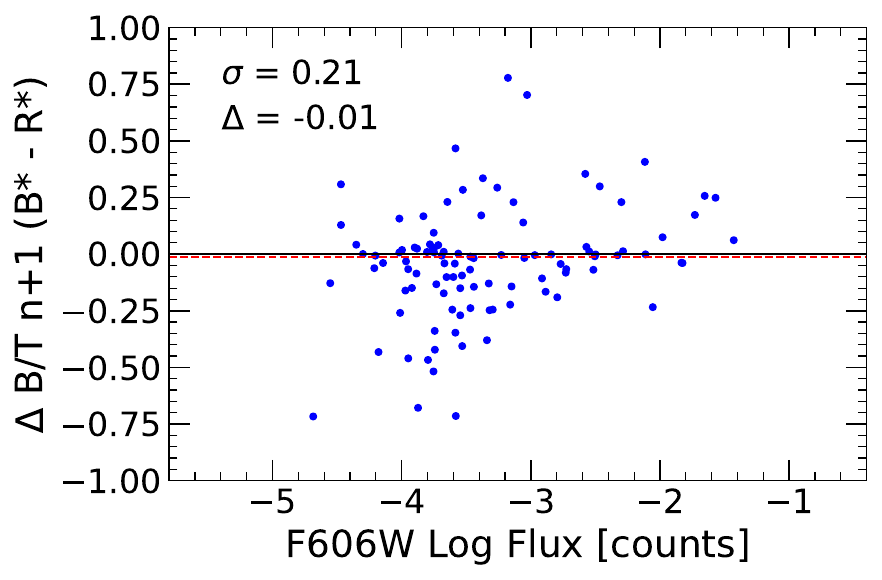}{0.3\textwidth}{}
          \fig{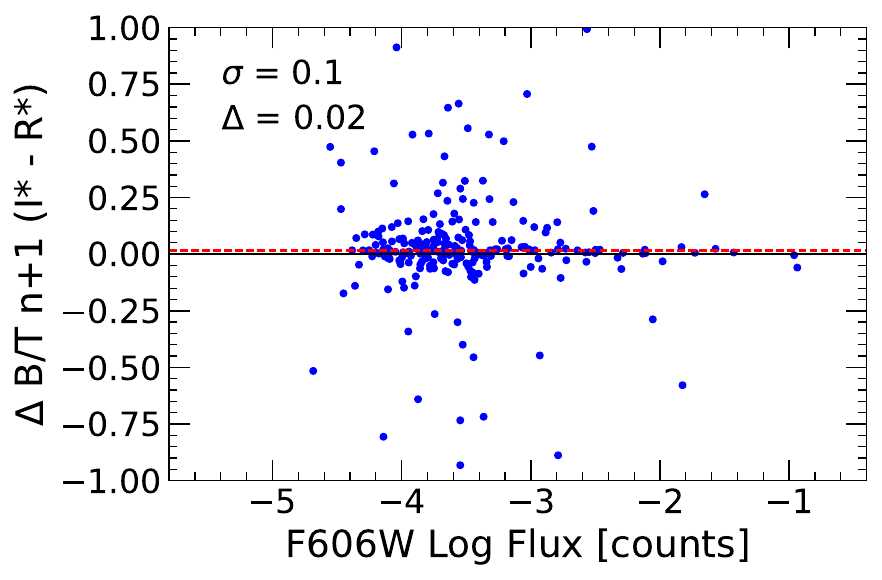}{0.3\textwidth}{}
          \fig{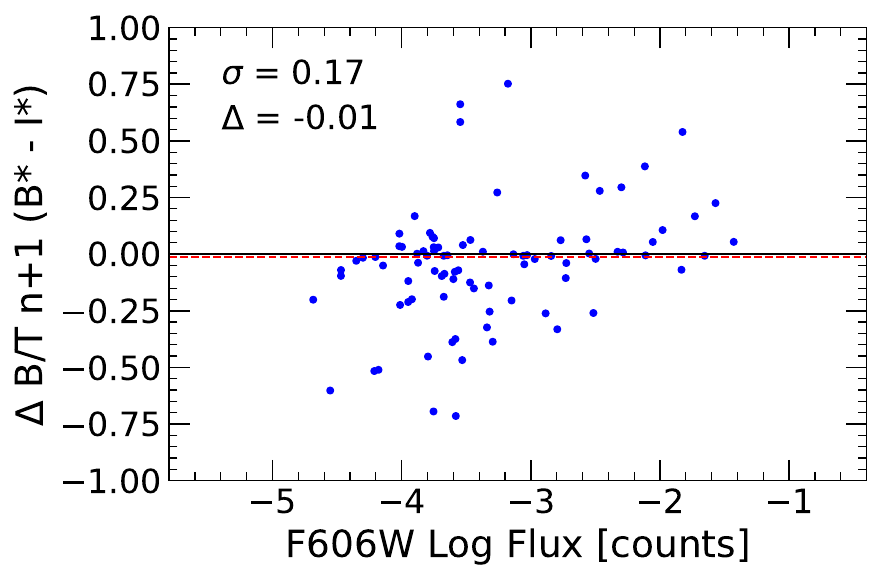}{0.3\textwidth}{}
          }
    \vspace*{-\baselineskip}
    \caption{Difference in HST-derived morphology metrics (log global $n$, log half-light radius ($R_{\text{eff}}$), log bulge $n$, B/T (4+1), B/T (n+1)) between images of the same galaxy widely separated in wavelength versus the single-component model flux (total photon counts) in the F606W band ($R^*$). The left column shows the differences between $B^*$ and $R^*$ properties, the middle column shows the differences between $I^*$ and $R^*$ properties, and the right column shows the differences between $B^*$ and $I^*$ properties. Medians and robust standard deviations (from 16th and 84th percentiles) are given as $\Delta$ and $\sigma$ respectively.}
    \label{kcorrection_plots}
\end{figure*}

\begin{figure}
    \centering
    \includegraphics[scale=0.6]{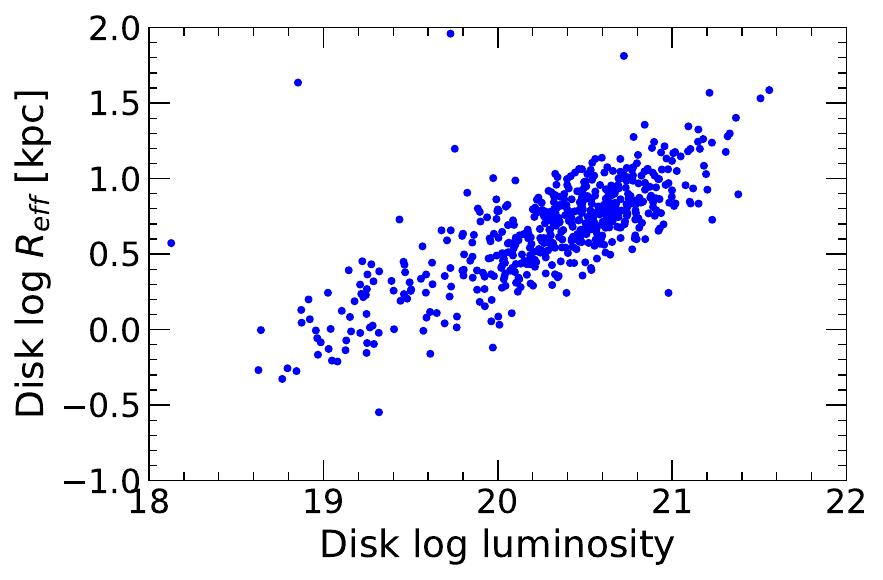}
    \caption{HST-derived size-luminosity relation for disk components of 569 face on galaxies (ellipticity $< 0.25$). The luminosity is in arbitrary units. Our method recovers the scaling relation between disk sizes and luminosities.}
    \label{disk_r50_relation_plot}
\end{figure}

\begin{figure*}
    \centering
    \gridline{\fig{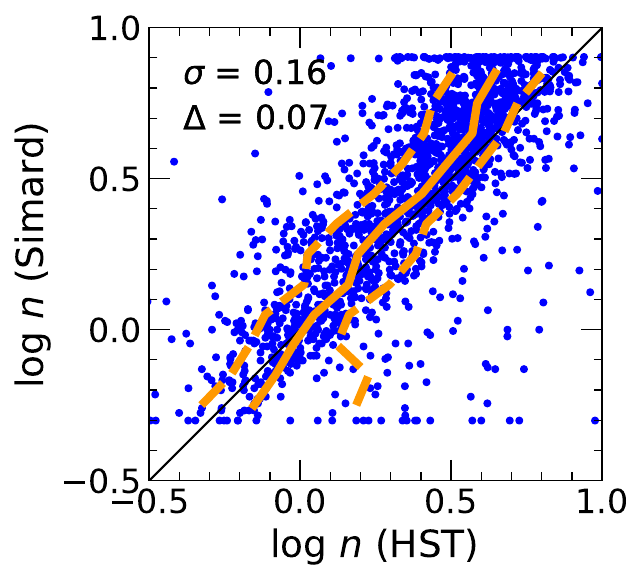}{0.3\textwidth}{}
          \fig{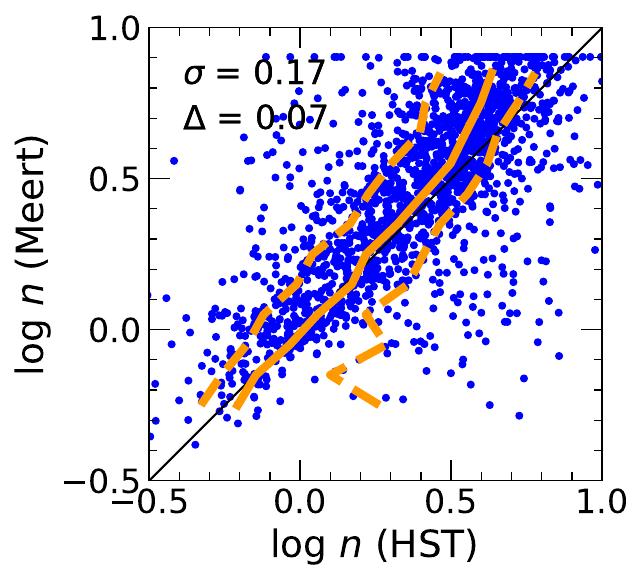}{0.3\textwidth}{}
          \fig{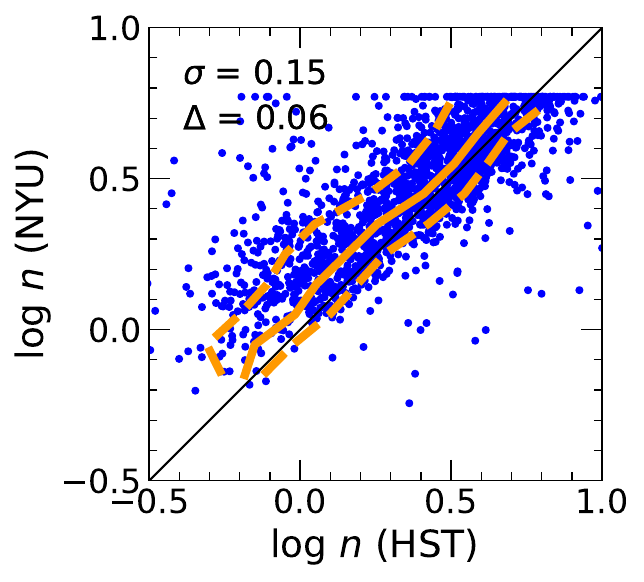}{0.3\textwidth}{}
          }
    \vspace*{-\baselineskip}
    \gridline{\fig{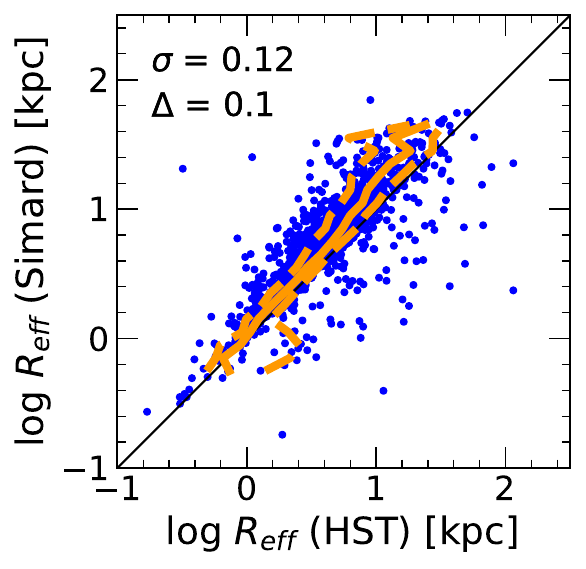}{0.2725\textwidth}{}
          \fig{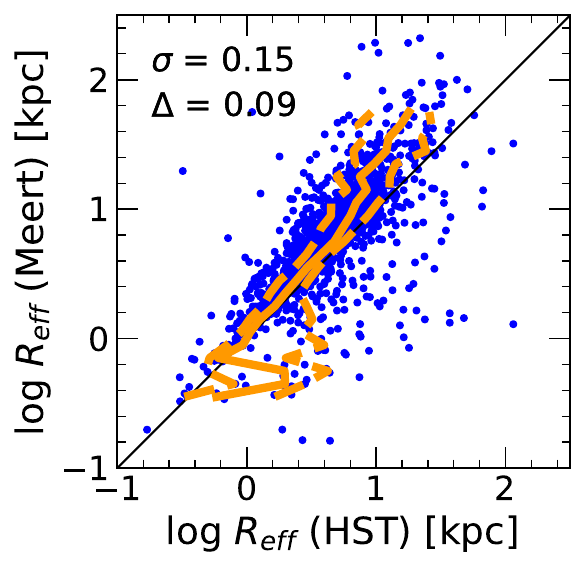}{0.2725\textwidth}{}
          \fig{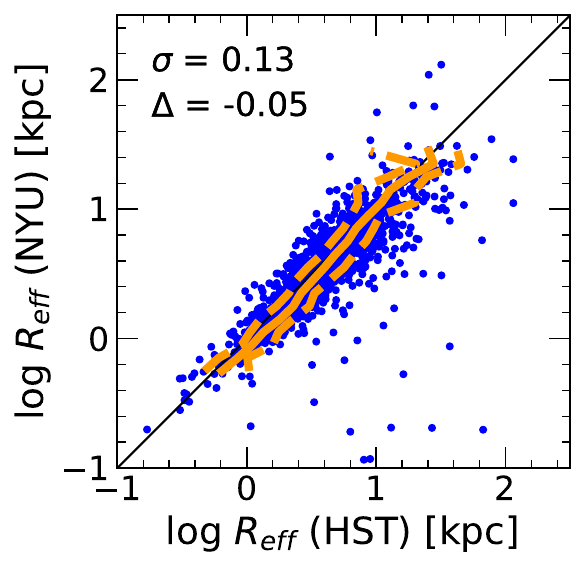}{0.2725\textwidth}{}
          }
    \vspace*{-\baselineskip}
    \caption{One-to-one comparison of HST to SDSS single S\'ersic fit parameters, log S\'ersic index ($n$) and log half-light radius ($R_{\text{eff}}$). The left column compares HST to the catalog of \citet{Simard2011BulgeDiskDecompCatalog}, the middle column compares to the \citet{Meert2015SDSSBulgeDiskDecompCatalog} catalog, and the right column compares to the NYU catalog of \citet{Blanton2005NYUcatalogForSDSS}. The robust standard deviation (from 16th and 84th percentiles; $\sigma$) and median offset ($\Delta$) of the SDSS (y-axis) minus the HST (x-axis) values are inset. Solid orange lines represent the running medians, while the dashed orange lines represent the 16th and 84th percentiles. Running medians are y-binned. Medians for bins with $<5$ objects are excluded for this figure and others of this type.}
    \label{simard_hst_1to1_comparison_plots}
\end{figure*}

\begin{figure*}
    \centering
    \gridline{\fig{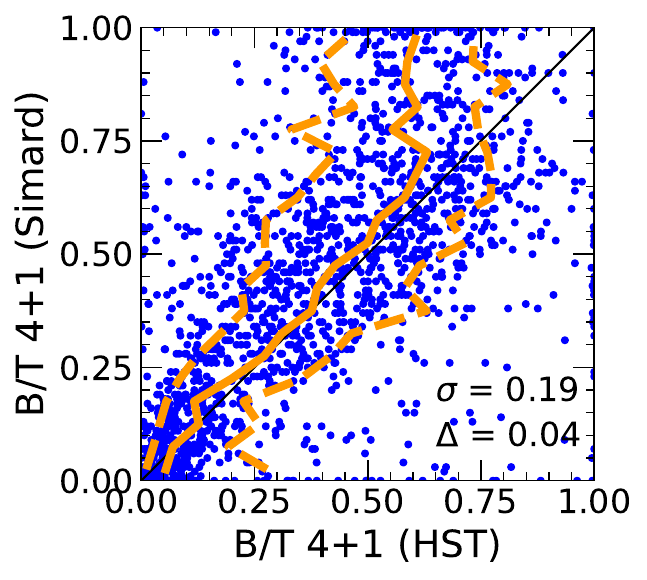}{0.35\textwidth}{}
          \fig{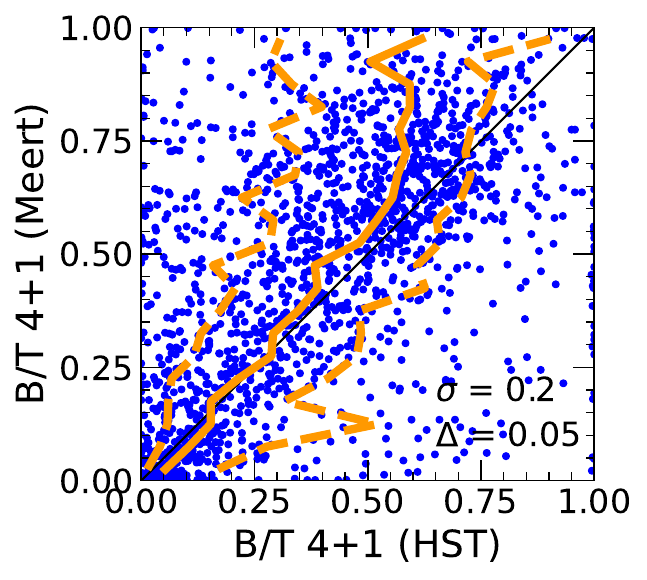}{0.35\textwidth}{}
          }
    \vspace*{-\baselineskip}
    \gridline{\fig{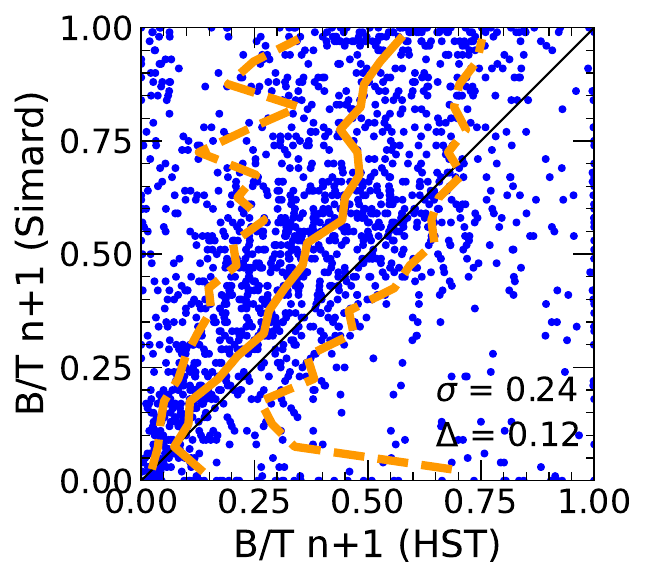}{0.35\textwidth}{}
          \fig{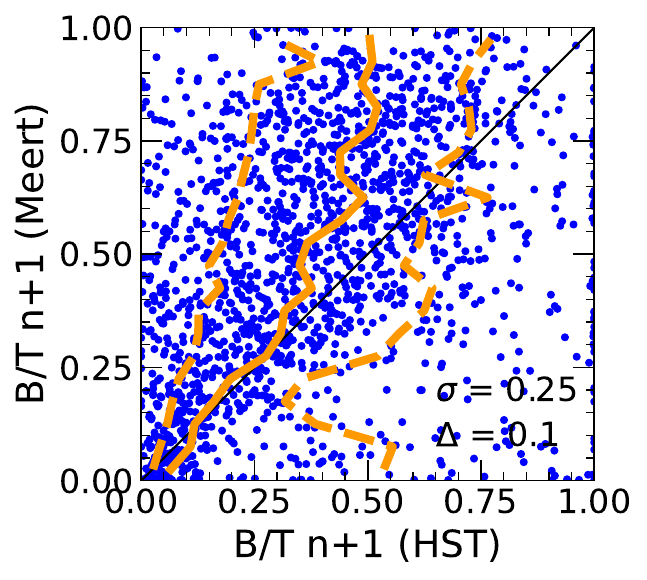}{0.35\textwidth}{}
          }
    \vspace*{-\baselineskip}
    \gridline{\fig{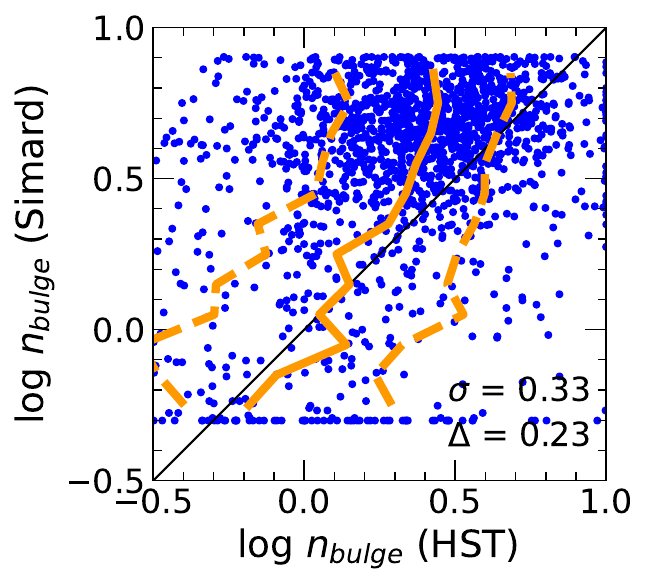}{0.35\textwidth}{}
          \fig{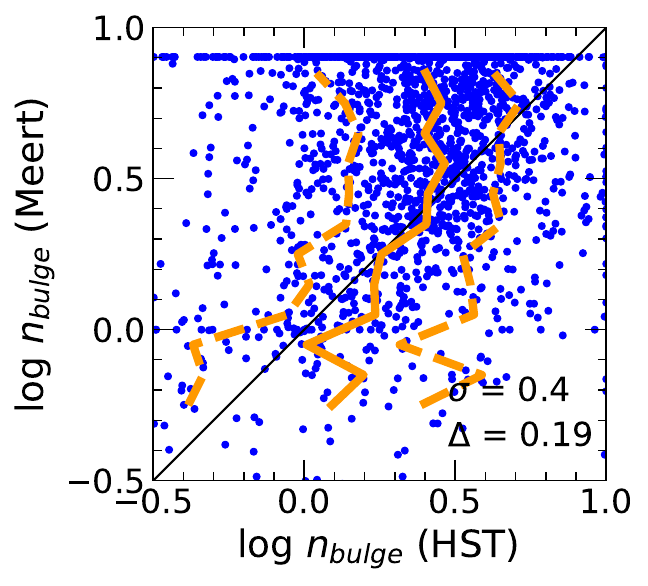}{0.35\textwidth}{}
          }
    \vspace*{-\baselineskip}
    \caption{One-to-one comparison of HST to SDSS two-component bulge/disk decompositions. Comparisons to \citet{Simard2011BulgeDiskDecompCatalog} are shown on the left and comparisons to \citet{Meert2015SDSSBulgeDiskDecompCatalog} are shown on the right. SDSS-derived parameters show significant bias for bulge-dominated galaxies. The robust standard deviation (from 16th and 84th percentiles; $\sigma$) and median offset ($\Delta$) of the SDSS (y-axis) minus the HST (x-axis) values are inset. Solid orange lines represent the y-binned running medians, while the dashed orange lines represent the 16th and 84th percentiles.}
    \label{simard_2comp_hst_1to1_comparison_plots}
\end{figure*}

\section{Assessment of HST-derived morphologies}
\label{Section:HSTAssessment}

In this section we evaluate the wavelength dependence and self-consistency of the morphologies obtained from the HST images. We also assess the disk size versus luminosity relation for the disks in our sample. For the wavelength assessments we consider 3 different wavelength regimes which we label as $B^*$, $R^*$, and $I^*$. $B^*$ refers to any of the F390W, F435W, F438W, F450W, and F475W filters. $R^*$ refers only to the F606W filter. $I^*$ refers to the F814W and F850LP filters. $B^*$ properties are taken from the bluest available filter in the $B^*$ filter set while $I^*$ properties are taken from the F814W image if available and the F850LP image otherwise. 

\subsection{Self-consistency of the HST-derived morphological parameters}
\label{Section:HSTAssessment:SelfConsistency}

We first consider the consistency of HST-derived morphologies for galaxies with multiple images at similar wavelengths. Figure \ref{BRI_self_consistency_plots} shows the difference in log $n$, log half-light radius ($R_{\text{eff}}$), log $n_{\text{bulge}}$, and B/T from both 4+1 and n+1 fits versus the model photon counts. The model counts used in Figure \ref{BRI_self_consistency_plots} are calculated as the total flux (in units of counts/second) from the single-component model fit multiplied by the exposure time associated with the image with smaller counts, and the difference is the parameter of the image with larger counts minus the parameter of the image with smaller counts. The flux on the x-axis is the lesser of the two model counts (in log scale) used to compute the difference on the y-axis. Showing the differences versus the flux in counts allows us to assess the impact of the constraining power of the images on the accuracy of the structural parameters, taking both the exposure time and apparent brightness into account. This is advantageous over using the exposure times or the fluxes in counts/sec alone; for example, a galaxy may have a short exposure time but is intrinsically bright and so has enough counts for good constraints on the fits anyway. On the other hand, an intrinsically dim galaxy with poor counting statistics may nonetheless have a long exposure time. 

We note that, while it is somewhat atypical to consider the logarithm of $n$ given the small dynamic range, we note that the character of $n$ is logarithmic; there is as much of a qualitative difference between $n = 1$ and $n = 2$ profiles as there is between $n = 4$ and $n = 8$ profiles. Consequently, the uncertainties in the logarithm of $n$ are much more even and readily interpreted than in linear $n$.

The median differences are consistent with zero for all properties in all 3 band groups, never exceeding 0.01. There is also a general trend across all properties and bands where the scatter decreases as the counts increase. Overall, we find no evidence of systematic differences in morphology among multiple images of the same galaxy. From the robust standard deviation (the difference in 16th and 84th percentile values) across all three band groups, we estimate the intrinsic precision of our HST measurements to be 0.05 dex (12$\%$) in $R_{\text{eff}}$, 0.07 dex (17$\%$) in $n$, 0.10 dex (26$\%$) in $n_{\text{bulge}}$ and 0.05 in B/T.  

While the median differences are consistent with zero, there are a number of outliers where the difference in morphology is significant despite the similarity in the wavelength of the images. For example, in the $R*$ $\Delta$ log $n_{\text{bulge}}$ panel (third row, middle column of Figure \ref{BRI_self_consistency_plots}) there are several galaxies with $\Delta$ log $n_{\text{bulge}} \sim 0.75$. The presence of these outliers is unexpected given that the same fitting routine is being applied uniformly to images of the same galaxies at similar wavelengths. Visual inspection of the most extreme outliers suggests that they often occur due to discrepancies in the image processing, such as a close neighbor which is deblended in one image but merged (i.e., part of the same segmentation map) with the target in the other image, which can bias the model fitting. 

We find that the scatter in parameters between and within band groups is not solely driven by a single property (e.g., exposure time), but is rather a consequence of the inhomogeneity of our HST dataset, to which differences in exposure time, wavelength, instruments, detectors, background, bias subtraction, and other similar factors all contribute.  Stochasticity in the fitting routine will also have a significant impact on the fits, since the fitting algorithm is a gradient descent method that may be vulnerable to local minima. The issue of local minima is exacerbated by the generally large number of pixels (typically on the order of thousands to tens of thousands) per image as well as the high resolution which produces more complex observed light profiles compared to SDSS. To fully resolve extreme outliers and improve the consistency of the fits is beyond the scope of this study, though we aim to make such improvements as part of future work. That being said, since we do not observe significant bulk offsets within or between the different band groups, the bulk trends with respect to SDSS are preserved.

\subsection{Wavelength dependence of the HST-derived morphological parameters}

Next we consider the comparison of HST-derived morphologies for galaxies with multiple images at different wavelengths. The goal is to determine whether there are systematic differences in the morphologies based on the observed wavelength. Figure \ref{kcorrection_plots} shows the difference in log global $n$, log half-light radius, log bulge $n$, and B/T from both 4+1 and n+1 fits versus the model flux (in total photon counts) from the F606W band ($R^*$) single component fit. In this context, the $R^*$ model flux serves as a fixed reference to assess the wavelength systematics as it is a measure common to each galaxy and is related to the constraining power of the $R^*$ fits. The left column of Figure \ref{kcorrection_plots} shows the difference between $B^*$ and $R^*$ properties, the middle column shows the difference between $I^*$ and $R^*$ properties, and the right column shows the difference between $B^*$ and $I^*$ properties.

The median offsets between any two band groups for all structural parameters are of order $0.01$ dex. Median S\'ersic indices are slightly smaller for $B^*$ compared to $R^*$ and $I^*$, as expected, but the difference is relatively small (around $10\%$). S\'ersic indices are essentially the same on average between the $R^*$ and $I^*$ bands. The robust standard deviations are roughly two times higher between $B^*$ and the other band groups compared to between $I^*$ and $R^*$. In all cases the standard deviation is greater than the estimated precision of the measurements from repeat measurements of the same galaxy in the same band group (Section \ref{Section:HSTAssessment:SelfConsistency}), from which we conclude that there are genuine differences in structural parameters at different wavelengths, but they do not constitute a bulk trend. Given the significantly higher standard deviations for $B^*$ bands and the small number of galaxies which only have $B^*$ images (27 galaxies), we exclude from the rest of the analysis the galaxies where we only have $B^*$ images. For the 1,716 remaining galaxies we adopt as nominal the properties from the image with highest model flux among $R^*$ and $I^*$ images.

\subsection{Size--luminosity relation for disks}

As a sanity-check on our two-component fits, we show in Figure \ref{disk_r50_relation_plot} the size-luminosity relation for disks in our sample. The disk radii and fluxes are taken from the 4+1 model fit. Only face-on disks with ellipticity $< 0.25$ are included in this assessment, which amounts to 569 galaxies. \citet{Simard2011BulgeDiskDecompCatalog} found that anomalies could be introduced in the disk size-luminosity space if the background estimation is poor. We find no evidence of such anomalies, suggesting that the disk model fits are robust. Like \citetalias{Simard2011BulgeDiskDecompCatalog}, we recover a clear scaling relationship between the disk size and the disk luminosity. 

\section{Assessment of systematics in SDSS-derived morphologies}
\label{Section:Results}

In this section we present comparisons between parameters determined in the literature from SDSS images and our measurements from HST images. All SDSS-derived properties are based on fits to SDSS \textit{r}-band imaging. 

\subsection{One-to-one comparison of SDSS-derived to HST-derived morphologies}

Figure \ref{simard_hst_1to1_comparison_plots} directly compares SDSS-derived single-component fit parameters (global Sersic index and effective radius) to our HST-derived single-component parameters. The left, middle, and right columns show comparisons to \citetalias{Simard2011BulgeDiskDecompCatalog}, \citetalias{Meert2015SDSSBulgeDiskDecompCatalog}), and NYU, respectively. In what follows we will assume HST-dervied parameters to be fiducial. We show running medians binned in SDSS-derived quantities, since that is the quantity that is available to a user. Note that the typical scatter ($\sim0.16$ dex for $n$ and $\sim0.13$ dex for $R_{\text{eff}}$) is significantly higher than the estimated precision of HST measurements (0.07 dex and 0.05 dex respectively) as well as the differences between $R^*$ and $I^*$, so the scatter essentially reflects the precision of the SDSS measurements. 

Both \citetalias{Simard2011BulgeDiskDecompCatalog} and \citetalias{Meert2015SDSSBulgeDiskDecompCatalog} tend to overpredict the global $n$ by as much as 0.2 dex when $n$ is relatively high, i.e., at log $n \gtrsim 0.5$, corresponding to bulge-dominated galaxies. The S\'ersic index from NYU does not show a significant systematic offset at high values of $n$, but this may be partially because the upper limit of $n$ for NYU is lower, artificially suppressing any upscatter. 

The simulations of \citep{Meert2013Simulations} also reveal systematics in the recovered global $n$ when fitting single-component profiles to intrinsically single-component galaxies, though their results differ from ours in that the systematics are opposite in sign; they find increasingly underpredicted global $n$ as the intrinsic $n$ increases. However, the majority of our sample are not strictly single-component and have some proportion of both a bulge and a disk component, which may account for the difference in our results. Regardless, it would seem that galaxies with high global $n$ are more vulnerable to systematic biases in morphology than those with lower global $n$. By increasing the resolution of their simulated images, \citep{Meert2013Simulations} are able to recover the global $n$ without bias for all galaxies, suggesting that the resolution may be primarily responsible for driving the bias in global $n$. 

NYU appears to predict systematically higher $n$ when HST $n$ is low (log $n \lesssim 0.4$) which is not seen for \citetalias{Meert2015SDSSBulgeDiskDecompCatalog} or \citetalias{Simard2011BulgeDiskDecompCatalog}. NYU fits assume an axisymmetric S\'ersic model, with the only free parameters being the amplitude, S\'ersic index, and effective radius, which may explain the differing character of the systematics with respect to the other SDSS fits. Indeed, we find that the difference in $R_{\text{eff}}$ between HST and NYU is strongly correlated with the galaxy ellipticity, such that the highest ellipticity galaxies have the largest differences in $R_{\text{eff}}$ (up to $\sim 0.3$ dex for ellipticity $\geq 0.75$). Therefore, the values of $R_{\text{eff}}$ from NYU are still subject to strong systematics, but these systematics do not dependent on the HST-derived $R_{\text{eff}}$. 

In contrast to NYU, the effective radius ($R_{\text{eff}}$; in kiloparsecs) from \citetalias{Simard2011BulgeDiskDecompCatalog} has a slight systematic offset to higher values compared to HST, with a median offset of 0.1 dex, and $R_{\text{eff}}$ from \citetalias{Meert2015SDSSBulgeDiskDecompCatalog} displays a similar trend. The offset appears to be driven primarily by galaxies with large (physical) sizes, as evidenced by the running median lines, for both \citetalias{Simard2011BulgeDiskDecompCatalog} and \citetalias{Meert2015SDSSBulgeDiskDecompCatalog}.  Notably, \citet{Meert2013Simulations} demonstrate that $R_{\text{eff}}$ may be overestimated when fitting a single-component model to an intrinsically two-component galaxy. In any case, $R_{\text{eff}}$ from \citetalias{Simard2011BulgeDiskDecompCatalog} appears to produce the least amount of bias and scatter overall.

Figure \ref{simard_2comp_hst_1to1_comparison_plots} compares two-component bulge/disk decomposition parameters from \citetalias{Simard2011BulgeDiskDecompCatalog} (left column) and \citetalias{Meert2015SDSSBulgeDiskDecompCatalog} (right column) to those from our HST bulge/disk decomposition. Overall, the correlation (scatter) between SDSS-derived and HST-derived two-component parameters is noticeably poorer than that of the single-component parameters. Of the two B/T ratios, the one determined using the 4+1 model is somewhat better correlated with HST than the n+1 model, but the ability to constrain B/T for any individual galaxy is still rather poor. B/T is reasonably well determined in the statistical sense; the B/T ratios from our 4+1 fits generally agree with SDSS, with only a 0.04 and 0.05 bulk offset with respect to \citetalias{Simard2011BulgeDiskDecompCatalog} and \citetalias{Meert2015SDSSBulgeDiskDecompCatalog}, respectively. However, like with global $n$, there is a systematic trend such that \citetalias{Simard2011BulgeDiskDecompCatalog} and \citetalias{Meert2015SDSSBulgeDiskDecompCatalog} both tend to overpredict B/T among high B/T galaxies, particularly those with HST B/T $\sim 0.6$. 

The values of $n_{\text{bulge}}$ from \citetalias{Simard2011BulgeDiskDecompCatalog} are essentially uncorrelated to the HST-derived values. Values of $n_{\text{bulge}}$ from \citetalias{Meert2015SDSSBulgeDiskDecompCatalog} are somewhat better correlated with the HST-derived values than \citetalias{Simard2011BulgeDiskDecompCatalog} especially when the $n_{\text{bulge}}$ from \citetalias{Meert2015SDSSBulgeDiskDecompCatalog} is relatively low, though the constraints for individual galaxies are rather weak like for B/T. \citetalias{Meert2015SDSSBulgeDiskDecompCatalog} also suffers from a `piling up' of galaxies with \citetalias{Meert2015SDSSBulgeDiskDecompCatalog} log $n_{\text{bulge}} = 0.9$, so the accuracy at a given HST-derived $n_{\text{bulge}}$ is poor for individual galaxies when using either catalog. Overall, we find that $n_{\text{bulge}}$ cannot be constrained from SDSS images. 

In conclusion, we do not find that any SDSS-based catalog performs considerably better than the other in all respects. We may prefer \citetalias{Meert2015SDSSBulgeDiskDecompCatalog} for S\'ersic indices of disky galaxies (log $n<0.4$) and NYU for S\'ersic indices of bulge-dominated galaxies (log $n>0.4$). On balance, NYU produces less biased $R_{\text{eff}}$ than the other two studies.

\begin{figure*}
    \centering
    \gridline{\fig{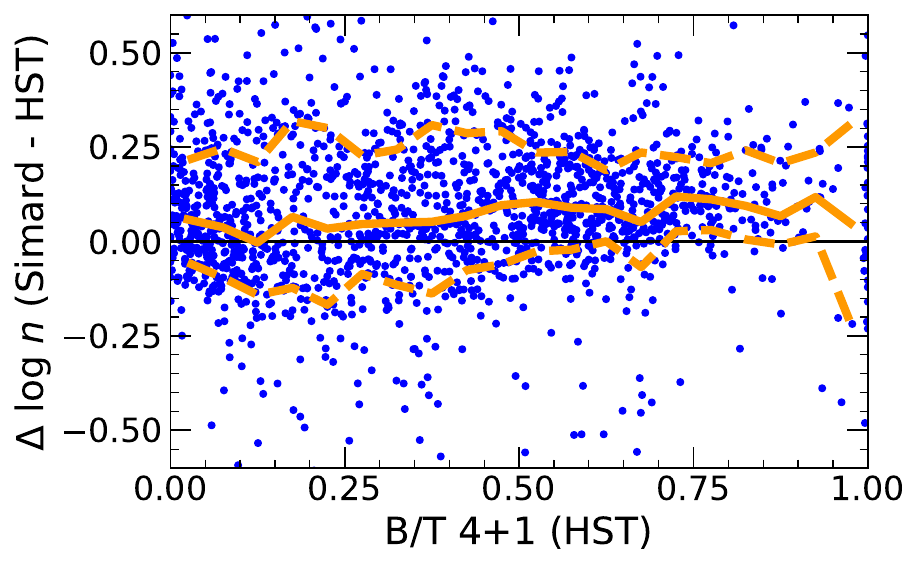}{0.4\textwidth}{}
          \fig{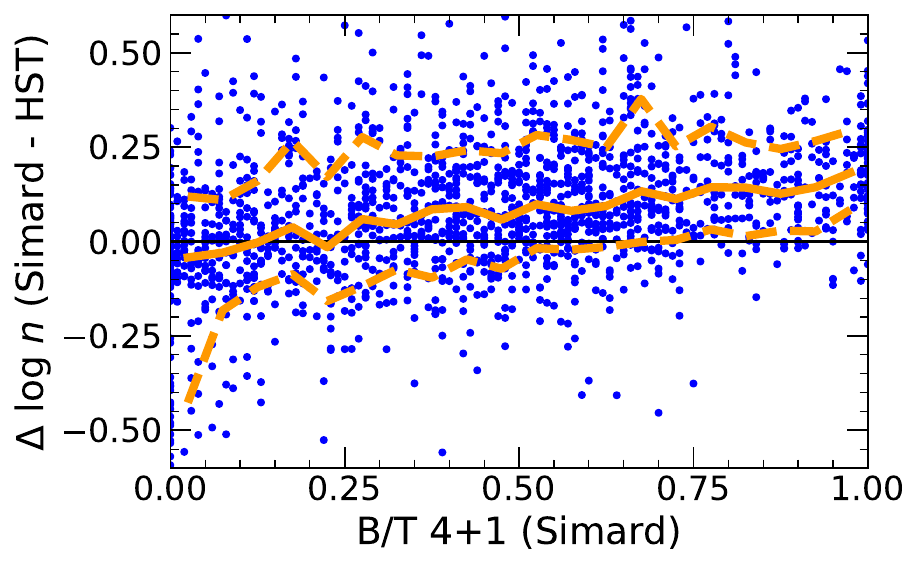}{0.4\textwidth}{}
          }
    \vspace*{-\baselineskip}
    \gridline{\fig{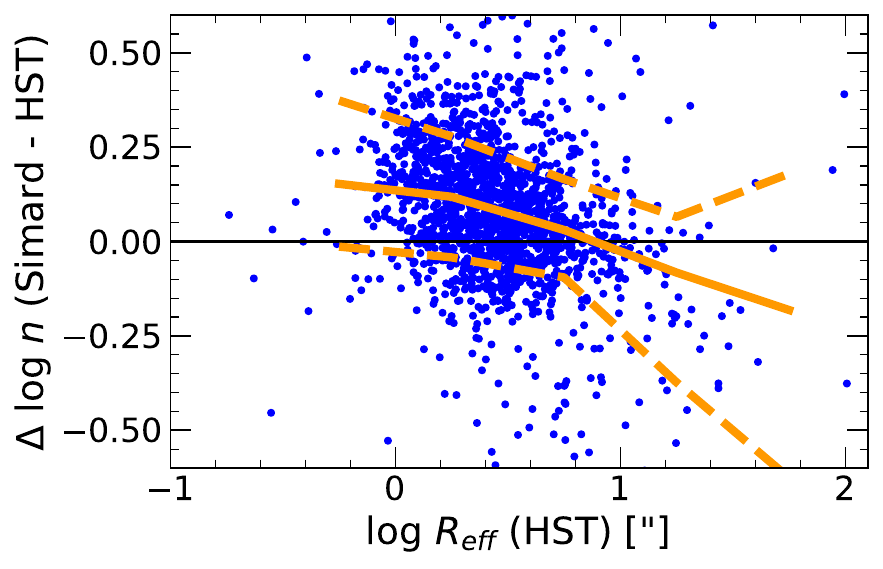}{0.4\textwidth}{}
          \fig{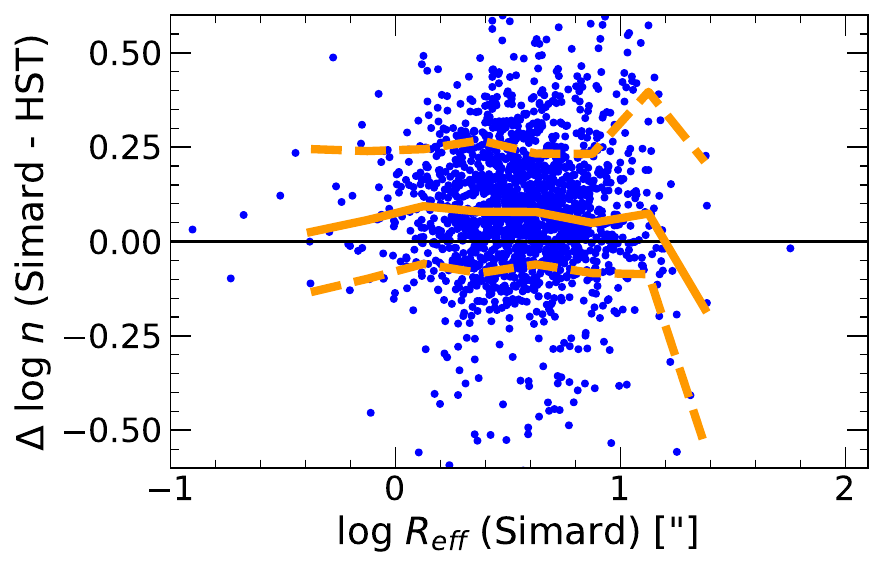}{0.4\textwidth}{}
          }
    \vspace*{-\baselineskip}
    \caption{Difference in global log $n$ between SDSS (\citetalias{Simard2011BulgeDiskDecompCatalog}) and HST versus the 4+1 B/T ratio for HST (top left) and SDSS (top right), and the angular half-light radius ($R_{\text{eff}}$) for HST (middle left) and SDSS (middle right). The systematic offset in log $n$ depends on both estimates of the B/T ratio and the HST angular size, but not the SDSS angular size. Solid orange lines represent the running medians while the dashed orange lines represent the 16th and 84th percentiles.}
    \label{morphology_versus_various_plots}
\end{figure*}

\subsection{Drivers of systematics between SDSS-derived and HST-derived S\'ersic indices}

We have thus far established that there are systematics in the SDSS-derived properties with respect to the HST-derived properties. In this section we seek to identify the regimes in which the systematics occur to ascertain whether they can be corrected. Overall, we have found the 2-component properties to be fundamentally poorly constrained while the global $R_{\text{eff}}$ is relatively robust, so in this section we focus on the systematics in the global S\'ersic index ($n$). We also focus only on \citetalias{Simard2011BulgeDiskDecompCatalog} since it is the most widely used of the three catalogs.

Figure \ref{morphology_versus_various_plots} shows the difference in global log $n$ between HST and \citetalias{Simard2011BulgeDiskDecompCatalog} versus 4+1 B/T and log effective radius. $\Delta$ log $n$ increases with both the HST-derived and SDSS-derived B/T, but the correlation is stronger with respect to the SDSS-derived B/T with higher peak median offset ($\sim 0.2$ dex as opposed to $\sim 0.1$ dex for HST-derived B/T) among the bulge-dominated galaxies. Disk-dominated galaxies (B/T $\sim 0$) do not appear to suffer from bulk systematics in global $n$. There is a strong correlation in $\Delta$ log $n$ with $R_{\text{eff}}$ from HST; the offset goes from $\sim 0.2$ dex at small radii to $\sim -0.1$ dex at large radii. No significant correlation is seen with respect to $R_{\text{eff}}$ from SDSS, though there is a bulk offset of $\sim 0.1$ dex. 

It is not obvious what property is driving the systematic trends with respect to size and B/T itself; is the difference in log $n$ driven by the galaxy size, the B/T ratio, or both? To test this, we recreated the panels in Figure \ref{morphology_versus_various_plots} using only galaxies in a small range of HST-derived sizes ($0.4 <$ log $R_{\text{eff}} < 0.6$) and, separately, only galaxies with HST-derived B/T in the range $0.4 < $ B/T $< 0.6$. The size-limited sample shows the same median trend in $\Delta$ log $n$ with respect to B/T, and the B/T-limited sample shows the same median trend in $R_{\text{eff}}$, showing that the two trends are entirely independent from one another. The results of \citet{Meert2013Simulations} suggest that image resolution can drive systematics in the estimation of global $n$, which may explain the systematics in the B/T ratio since it is correlated with global $n$. 

\begin{figure}
    \centering
    \includegraphics[scale=0.75]{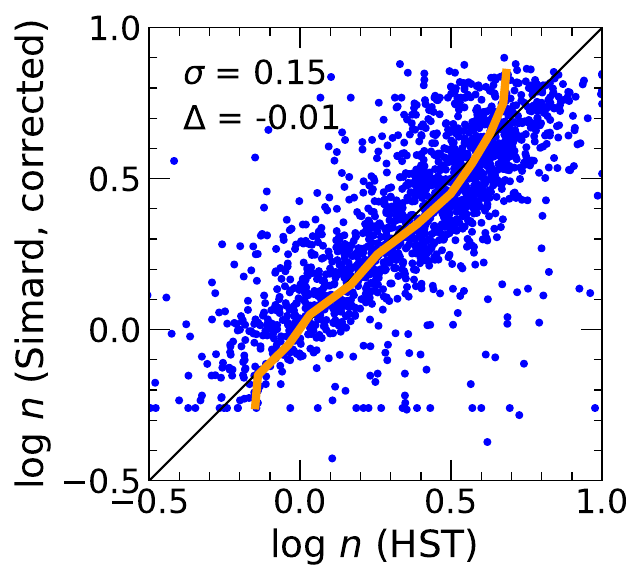}
    \caption{One-to-one comparison of global log $n$ between SDSS (\citetalias{Simard2011BulgeDiskDecompCatalog}) and HST where the SDSS-derived estimate has been empirically corrected using a line fit to the data in the top right panel in Figure \ref{morphology_versus_various_plots}.}
    \label{corrected_totaln_plots}
\end{figure}

Of the two effects, we can correct only the systematics correlated with the structural parameter (B/T) itself. The correction is obtained by fitting a line to $\Delta$log $n$ versus 4+1 B/T from \citetalias{Simard2011BulgeDiskDecompCatalog}:


$$ \text{ log } n \text{ (HST)} = \text{ log } n \text{ (S11)} - 0.261 \times \text{B/T} + 0.042 $$

Figure \ref{corrected_totaln_plots} shows the corrected one-to-one global log $n$ comparison (i.e., the corrected version of the top left panel of Figure \ref{simard_hst_1to1_comparison_plots}). The agreement in global log $n$ is significantly improved, in particular for the log $n \gtrsim 0.5$ galaxies. While in principle we could have constructed a correction as a function of $n$ (from \citetalias{Simard2011BulgeDiskDecompCatalog}) itself, by using B/T instead we avoid issues with correlated values (i.e., the correction to $n$ depending on $n$ itself) and also more effectively break up the pile up at the extremes of the $n$ distribution (Figure \ref{simard_hst_1to1_comparison_plots}).

\section{Conclusions}
\label{Section:Conclusions}

In this work we have used of a sample of 1,743 SDSS galaxies with serendipitous HST imaging to assess the reliability of morphological properties obtained from ground-based optical imaging of galaxies at $0.01 < z < 0.3$. We fit three types of profiles to the HST images in a manner similar to \citet{Simard2011BulgeDiskDecompCatalog}: 1) a single S\'ersic profile, a two-component $n = 4$ S\'ersic bulge plus an exponential disk (4+1), and a two-component free S\'ersic bulge plus an exponential disk (n+1). We compare HST measurements of the single-component S\'ersic $n$ (global $n$), single-component effective radius ($R_{\text{eff}}$), bulge S\'ersic index ($n_{\text{bulge}}$), and bulge-to-total ratio (B/T) from both the 4+1 and n+1 fits to the same measurements obtained from SDSS \textit{r}-band imaging of the same galaxies, in particular those of \citet[][S11]{Simard2011BulgeDiskDecompCatalog}, but also of \citet[][M15]{Meert2015SDSSBulgeDiskDecompCatalog} and \citet[][NYU]{Blanton2005NYUcatalogForSDSS}. We also evaluated the agreement between HST-derived and SDSS-derived morphologies in the context of other properties to determine in which regimes the differences are most significant. Our main conclusions are as follows:

\begin{enumerate}

\item We find no systematic trends of HST-derived global Sersic indices ($n$) as a function of wavelength within the optical range. However, there do exist real differences that contribute to the scatter between $n$ measured in different bands. 

\item Typical measurement errors for HST-derived quantities are 0.05 dex ($12\%$) in $R_{\text{eff}}$, 0.07 dex ($17\%$) in global $n$, 0.10 dex ($26\%$) in $n_{\text{bulge}}$, and 0.05 in B/T. The measurement errors for SDSS-derived (S11) quantities are 0.12 dex ($32\%$) in $R_{\text{eff}}$, 0.16 dex ($45\%$) in global $n$, 0.33 ($113\%$) in $n_{\text{bulge}}$, and 0.19 in B/T. 

\item Of the three SDSS catalogs considered, none perform significantly better than the others overall. Global S\'ersic indices from \citet{Meert2015SDSSBulgeDiskDecompCatalog} may be preferred for the lack of bias for galaxies with log $n<2.5$. NYU S\'ersic indices may be preferred for galaxies with log $n>2.5$; S11 and M15 systematically overestimate global $n$ for log $n>2.5$ galaxies by as much as $20\%$. S11 and M15 systematically overestimate $R_{\text{eff}}$ by $\sim 0.1$ dex for the largest galaxies (log $R_{\text{eff}} \gtrsim 0.6$). NYU estimates of $R_{\text{eff}}$ show no systematics with respect to the HST-derived $R_{\text{eff}}$, but become increasingly biased as galaxy ellipticity increases (up to $\sim 0.3$ dex at ellipticity $\geq 0.75$). $R_{\text{eff}}$ from S11 are preferred overall. 

\item In general, SDSS-derived single-component parameters (global $n$, $R_{\text{eff}}$) are better constrained than the SDSS-derived two-component parameters.

\item Measurements of $n_{\text{bulge}}$ from SDSS are quite unreliable. 

\item Measurements of the B/T ratio from SDSS using fixed bulge ($n_{\text{bulge}} = 4$) plus disk models are unbiased for galaxies with B/T $\lesssim 0.6$, albeit with large scatter for individual galaxies. Galaxies with higher B/T tend to have significantly overpredicted B/T ratios compared to HST.

\item An empirical correction to the SDSS global log $n$ based on the SDSS-derived B/T ratio is able to account for most of the systematic differences in global $n$ between SDSS and HST. The correction is estimated based on S11 but would be similar for M15.

\end{enumerate}

This research made use of {\tt Photutils}, an Astropy package for
detection and photometry of astronomical sources \citep{PHOTUTILS_larry_bradley_2022_6825092}. This research also made use of {\tt Petrofit} \citep{PETROFIT_Geda_2022}, a
package based on {\tt Photutils}, for calculating Petrosian properties
and fitting galaxy light profiles.

This work was supported through NASA award 80NSSC20K0440. 

Based on observations made with the NASA/ESA Hubble Space Telescope, and obtained from the Hubble Legacy Archive, which is a collaboration between the Space Telescope Science Institute (STScI/NASA), the Space Telescope European Coordinating Facility (ST-ECF/ESAC/ESA) and the Canadian Astronomy Data Centre (CADC/NRC/CSA).

\bibliography{Bibliography}
\bibliographystyle{aasjournal}

\end{document}